\documentclass[compsoc, conference, a4paper, 10pt, times]{IEEEtran}

\usepackage{cite}
\usepackage{amsmath,amssymb,amsfonts}
\usepackage{algorithmic}
\usepackage{graphicx}
\usepackage{textcomp}
\usepackage{xcolor}
\usepackage{booktabs}
\usepackage[hidelinks]{hyperref}

\usepackage{times}
\usepackage{subfiles} %
\usepackage{array}
\usepackage{tabularx}
\usepackage{longtable}
\usepackage{caption}
\usepackage{subcaption}
\usepackage{layouts}
\usepackage{pgf}
\usepackage{cleveref}
\usepackage{adjustbox}
\usepackage{bbm}
\usepackage{relsize}
\usepackage{enumitem}

\begin{document}

\title{ATLAS: Automatically Detecting Discrepancies \\ Between Privacy Policies and Privacy Labels}

\makeatletter
\newcommand{\linebreakand}{%
  \end{@IEEEauthorhalign}
  \hfill\mbox{}\par
  \mbox{}\hfill\begin{@IEEEauthorhalign}
}
\makeatother

\author{
    \IEEEauthorblockN{Akshath Jain\IEEEauthorrefmark{1}, David Rodriguez\IEEEauthorrefmark{2}, Jose M. del Alamo\IEEEauthorrefmark{2}, and Norman Sadeh\IEEEauthorrefmark{1}}\\
    \IEEEauthorblockA{\IEEEauthorrefmark{1}School of Computer Science, Carnegie Mellon University
    \\arjain@andrew.cmu.edu, sadeh@cs.cmu.edu}\\
    \IEEEauthorblockA{\IEEEauthorrefmark{2}Universidad Politécnica de Madrid
    \\\{david.rtorrado, jm.delalamo\}@upm.es}
}

\maketitle

\begin{abstract}
Privacy policies are long, complex documents that end-users seldom read. Privacy labels aim to ameliorate these issues by providing succinct summaries of salient data practices. In December 2020, Apple began requiring that app developers submit privacy labels describing their apps' data practices. Yet, research suggests that app developers often struggle to do so. In this paper, we automatically identify possible discrepancies between mobile app privacy policies and their privacy labels. Such discrepancies could be indicators of potential privacy compliance issues.

We introduce the Automated Privacy Label Analysis System (ATLAS). ATLAS includes three components: a pipeline to systematically retrieve iOS App Store listings and privacy policies; an ensemble-based classifier capable of predicting privacy labels from the text of privacy policies with 91.3\% accuracy using state-of-the-art NLP techniques; and a discrepancy analysis mechanism that enables a large-scale privacy analysis of the iOS App Store.

Our system has enabled us to analyze 354,725 iOS apps. We find several interesting trends. For example, only 40.3\% of apps in the App Store provide easily accessible privacy policies, and only 29.6\% of apps provide both accessible privacy policies and privacy labels. Among apps that provide both, 88.0\% have at least one possible discrepancy between the text of their privacy policy and their privacy label, which could be indicative of a potential compliance issue. We find that, on average, apps have 5.32 such potential compliance issues.

We hope that ATLAS will help app developers, researchers, regulators, and mobile app stores alike. For example, app developers could use our classifier to check for discrepancies between their privacy policies and privacy labels, and regulators could use our system to help review apps at scale for potential compliance issues.
\end{abstract}


\section{Introduction}

Notice is a cornerstone of privacy: entities collecting information are expected to disclose the types of data collected, and how it is used. Privacy policies serve as the primary mechanism for notice, yet research has shown that privacy policies are long, complex documents that users seldom read \cite{mcdonald2008cost} \cite{sadeh2013usable} \cite{reidenberg2015disagreeable}.

Privacy labels aim to ameliorate this issue by providing succinct descriptions of salient data practices in an easy to consume format. In December 2020, Apple began requiring that developers include privacy labels for the apps they publish on the iOS App Store. However, recent research suggests that mobile app developers often struggle to understand and disclose their data practices. \cite{li2022understanding} \cite{zhang2023privacy}.

In this paper, we present work that addresses three sets of research questions concerning the adoption and content of privacy policies and privacy labels in the United States iOS App Store.

\begin{enumerate}[leftmargin=*]
    \item[1.] What is the state of privacy policy and privacy label adoption among iOS apps? Can app privacy policies be easily accessed -- specifically, how many apps have direct links to their privacy policies in the App Store? What percentage of apps have privacy labels? What percentage have both?
    \item[2.] Is it possible to predict privacy labels from the text of privacy policies? Prior research indicates that developers struggle to create accurate labels. As a result, privacy labels may not always reflect actual data practices \cite{li2022understanding}, \cite{koch2022keeping}. So, despite privacy labels being noisy, is it possible to train reliable classifiers?
    \item[3.] Finally, are privacy labels consistent with the text of privacy policies? What types of discrepancies occur between privacy policies and labels? How many discrepancies do apps have on average? Is there a correlation between these rates and the popularity of mobile apps?
\end{enumerate}

To enable us to answer these research questions, we developed the Automated Privacy Label Analysis System (ATLAS). Our system analyzed iOS app metadata, privacy policies, and privacy labels to flag instances where privacy labels were not consistent with the text of their privacy policies, which we characterized as potential compliance issues. This was done using state-of-the-art natural language processing techniques. ATLAS was designed to be highly scalable and has enabled us to analyze 354,725 iOS apps. This paper makes several key contributions:

\begin{enumerate}[leftmargin=*]
    \item[1.] A scalable pipeline that enables systematically scraping iOS metadata, including privacy policies URLs and privacy labels. We include a machine learning model that determines if an app's privacy policy URL actually leads to an English-language privacy policy.
    \item[2.] An ensemble-based classifier that can effectively generate privacy labels from the text of privacy policies. This enabled our classifier to identify if a privacy policy ``Collects'' or ``Does Not Collect'' a data type, for 32 separate data types.
    \item[3.] A privacy analysis of the iOS App Store. We provided an extensive analysis of data practice disclosure discrepancies between the text of privacy policies and their corresponding privacy labels in the iOS App Store. We also include metrics such as privacy policy accessibility, privacy label adoption, and potential compliance issue existence.
\end{enumerate}

We hope that ATLAS will help app developers, researchers, regulators, and mobile app stores alike. For example, app developers could use our classifier to check for discrepancies between their privacy policies and privacy labels. Meanwhile, app store operators and regulators could use our system to monitor discrepancy trends to effectively focus efforts on apps likely to have potential compliance issues.

\section{Background and Related Work}

\subsection{Privacy Labels}

\subsubsection{iOS Privacy Labels}\label{sec:ios_privacy_labels}

\begin{figure}[h!]
    \centering
    \includegraphics[width=0.35\linewidth]{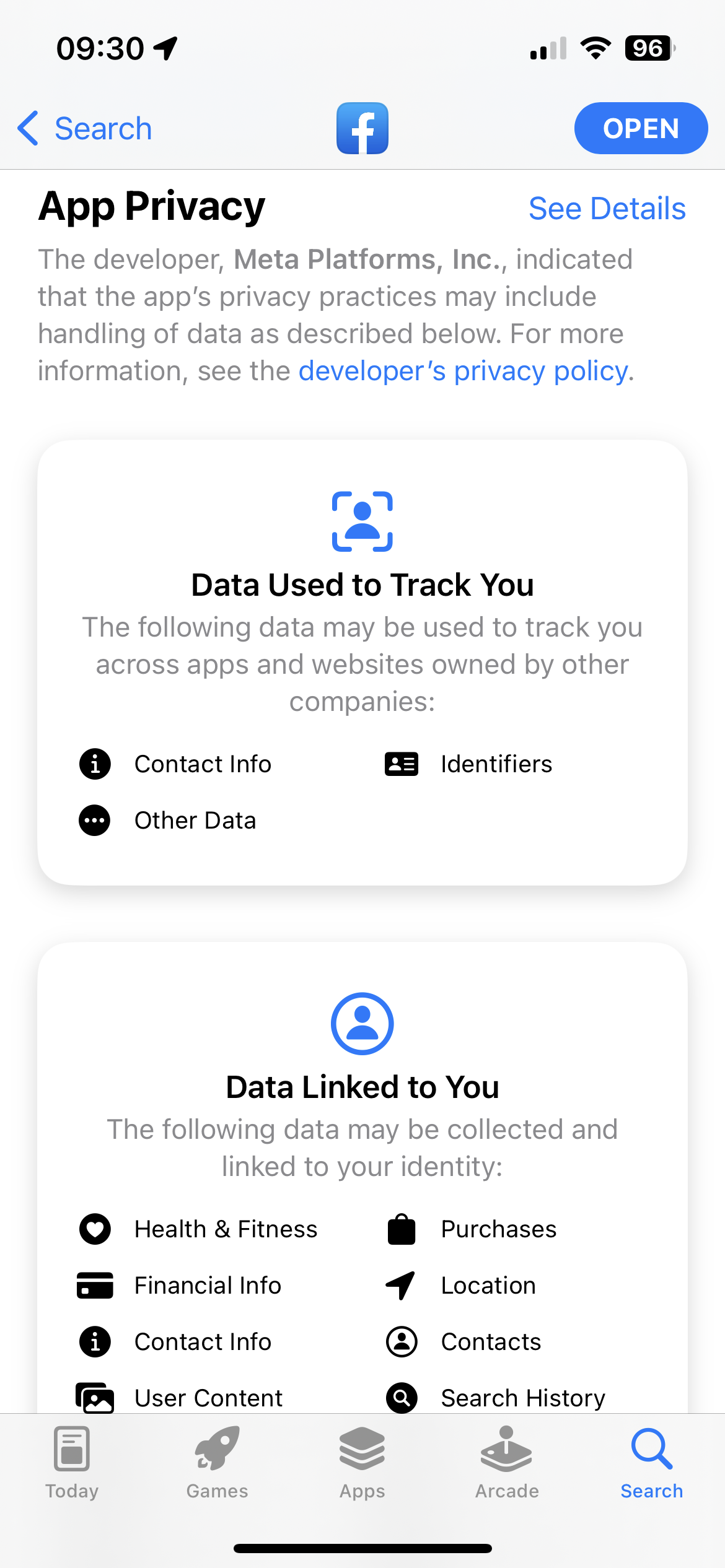}
    \includegraphics[width=0.35\linewidth]{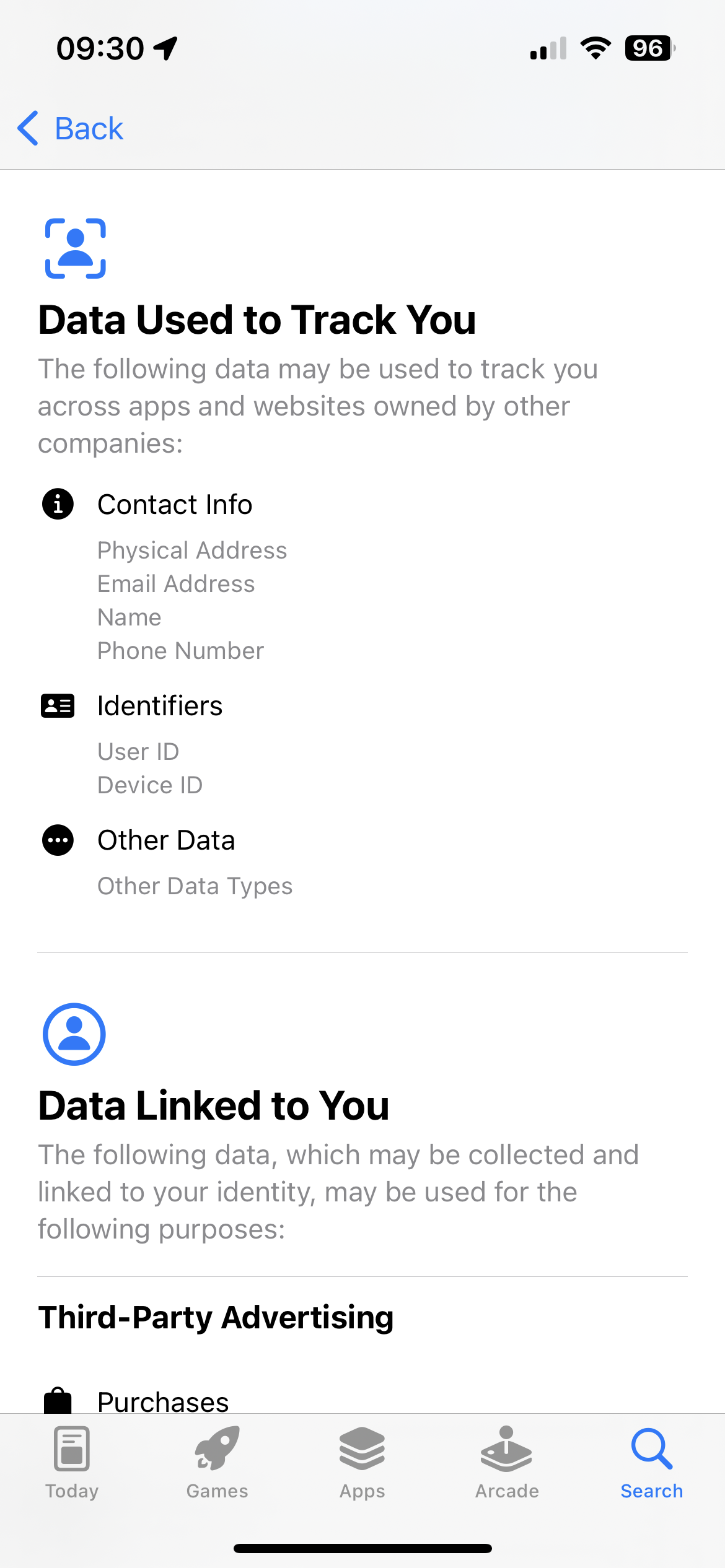}
    \caption[Example iOS Privacy Label]{An example of a privacy label within the iOS App Store. The image on the left shows an overview of the app's privacy label. The image on the right depicts a more detailed summary of the app's data collection.}
    \label{fig:privacy-label}
\end{figure}

In December 2020, Apple began requiring that all developers include ``App Privacy Details'' describing their apps' data collection practices when uploading new versions of their apps, arguably the largest adoption of privacy labels to date \cite{apple2020labels}. %
The implementation is largely similar to the model proposed by researchers in 2013 \cite{kelley2013privacy}.

Figure \ref{fig:privacy-label} shows an example of an iOS privacy label available within the iOS App Store. Apple takes a multilayered approach to privacy labels: users first encounter a summarized version, but can choose to ``See Details,'' which provides a more comprehensive overview of the app's data collection practices. %

The iOS privacy label consists of four parts. First, developers declare the data type(s) being collected by their app. Then, for each data type, developers are required to report how the data type is being used. Next, developers must specify if the data type is ``Linked to You,'' which indicates that it is being collected non-anonymously. Finally, if the developer declares that the data type is ``Linked to You,'' then they must declare if it is ``Used to Track You,'' which indicates linking collected data for targeted advertising or third-party data sharing \cite{apple2020labels}. %

\subsubsection{Issues with iOS Privacy Labels}

While iOS privacy labels are intended to help inform end-users about apps' data practices, they are not without issue. Recent research suggests that developers face challenges in creating accurate privacy labels \cite{li2022understanding} \cite{gardner2022helping}. After conducting interviews with iOS developers, the authors found that misconceptions about privacy labels often resulted in under- or over-reporting data collection \cite{li2022understanding}. Koch et al. conducted an ``exploratory statistical evaluation'' of 11,074 iOS apps, and found that only a ``small number of apps provide privacy labels'' \cite{koch2022keeping}. They also conducted a dynamic analysis of a small subset of 1,687 apps, finding that 276 (16\%) violated their own privacy labels by transmitting data without declaration \cite{koch2022keeping}. This work expands on Koch et al. by conducting a broader statistical evaluation on a much larger dataset, and by investigating to what extent privacy labels are inconsistent with the text of their privacy policies.

\subsubsection{Generating iOS Privacy Labels}

While iOS privacy labels have several issues, Li et al. suggest that auto-generating privacy labels might help increase their accuracy \cite{li2022understanding}. Tools, such as Privacy Label Wiz, aid developers in creating privacy labels for iOS apps by statically analyzing iOS app source code for signatures such as plist permission strings, import statements, and class instantiations \cite{gardner2022helping}. %
Importantly, these tools can help developers comply with regulations, such as GDPR, which require accurate privacy notices \cite{zimmeck2019compliance}.

Whereas Gardner et al. only focused on a small subset of iOS privacy labels (9 of 32), this work aims to analyze data collection disclosure for all 32 data types \cite{gardner2022helping}. Moreover, while \cite{zimmeck2021privacyflash} and \cite{gardner2022helping} used static code analysis, this work uses state-of-the-art natural language processing techniques to generate privacy labels directly from the text of privacy policies and compare the predicted labels with those reported within the App Store.

\subsection{Applicable Legislation}

Regulatory requirements for privacy disclosures vary by jurisdiction. However, the state of California and the European Union -- two of the largest digital markets -- have stringent privacy disclosure requirements. California requires compliance through the CCPA and the European Union through GDPR. Failure to provide accurate privacy disclosures in both jurisdictions could have significant legal ramifications.

\subsection{Automated Mobile App Privacy Analyses}

With millions of mobile apps, automation is the only way to analyze privacy practices at scale. Dynamic analysis systems, such as TaintDroid, actively run apps and monitor their behavior \cite{enck2014taintdroid} \cite{reardon201950}. While these techniques give a true representation of an app's behavior, it has significant overhead, limiting scale. Dynamic analysis systems have also largely been limited to Android, though recent systems have focused on iOS as well \cite{xiao2022lalaine}. Static analysis systems are far more scalable -- by reasoning about source code, systems like MAPS have been able to analyze over a million Android apps \cite{zimmeck2017automated} \cite{zimmeck2019maps}.

More recent work has focused on analyzing mobile app privacy on iOS. Kollnig et al. found that Apple's recent changes requiring user permission to access device identifiers and developers include privacy labels has made tracking more difficult \cite{kollnig2022goodbye}. Balash et al. conducted a large-scale longitudinal analysis of the App Store over 36 weeks, finding that of 1.6 million apps, only 60.5\% of apps provided privacy labels \cite{balash2022longitudinal}. Xiao et al. conducted a small scale privacy analysis by comparing the privacy labels and binaries of 5,102 iOS apps, finding many instances of non-compliance \cite{xiao2022lalaine}.

\subsection{Natural Language Processing Techniques for Document Classification}

Natural Language Processing (NLP) has been a cornerstone in semantically understanding and parsing privacy policies. Automated compliance systems, such as  MAPS and ATLAS, require the use of NLP to analyze hundreds of thousands of privacy policies that would otherwise take human annotators years to accomplish. Past research has detailed the effectiveness of using NLP techniques to analyze privacy policies for compliance analysis using high-quality annotator labeled data (i.e. annotations per privacy policy segment) from the APP-350 corpus \cite{story2019natural}. The authors formulated the identification of privacy practice statements as a classification problem using one classifier per privacy practice. \cite{story2019natural}.

This paper replicates the approach, with several key differences. First, instead of focusing on policy segments, this work formulates identification of data collection as a document classification problem. Second, instead of using annotator labeled data (such as from the APP-350 corpus), we use developer reported iOS privacy labels as ground truth to describe the text of privacy policies. Unlike annotator labeled data, privacy labels may not be consistent with the text of privacy policies because developers struggle to accurately create them \cite{gardner2022helping} \cite{li2022understanding}. Finally, we experiment with additional state-of-the-art model architectures for document classification.

Document classification is an extensively researched field \cite{kim2014convolutional} \cite{zhang2015character} \cite{yang2016hierarchical} \cite{liu2017deep}. Current state-of-the-art techniques are based on the transformer model (such as BERT and RoBERTa), which sidesteps traditional methods such as recurrence \cite{vaswani2017attention} \cite{devlin2018bert} \cite{liu2019roberta}. However, documents are typically longer than the maximum sequence length allowed by models such as BERT and RoBERTa -- 512 tokens. The Longformer model increases input sequence length to 4096 tokens, which is better suited for document classification \cite{beltagy2020longformer}. However, transformer-based models have large computational overhead. The RegLSTM model was proposed as a lightweight alternative for document classification and has been shown to outperform transformers \cite{adhikari2019rethinking}.

\section{A Distributed Pipeline for Automated Analysis}
\label{chap:3-scaling}

\subsection{iOS App Sampling Strategy}

We first began by identifying a candidate list of apps to analyze. Prior work has relied on crawling mobile app stores for app discovery; however, we found that not to be necessary for this work \cite{zimmeck2019maps}. Fortunately, Apple publishes a categorized, alphabetical list of all available iOS Apps, including popular apps per category \cite{apple2023catalog}. On January 29$^{\text{th}}$, 2023, we systematically crawled and scraped the website to assemble a list of 918,293 unique apps available on the United States iOS App Store. Of those apps, 4,846 are classified as popular apps.

Next, we devised a sampling strategy to pick a subset of apps to analyze. Conducting a simple random sample of the entire App Store is the easiest way to generate a representative sample of the entire App Store; however, this approach is likely to miss heavy-hitters: frequently downloaded apps that are more likely to be present on users's devices -- popular apps. Conversely, sampling only popular apps leads to a biased representation of the App Store, as many apps are missed. We devised a hybrid sampling strategy to create a set of apps likely to be present on user's devices \textit{and} an unbiased representation of apps available on the iOS App Store: we sampled all 4,846 popular apps in addition to a randomly selected set of 350,000 non-popular apps. In total, our final dataset comprised of 354,725 apps, as some app listings were unable to be loaded.

\subsection{Identifying Privacy Policies}

iOS apps are required to provide a URL to their privacy policy. However, in many cases, these URLs would lead to landing pages, or other unrelated webpages. To accurately obtain privacy policies, we developed a logistic regression classifier to determine if pages were English-language privacy policies, similar to prior work \cite{zimmeck2019maps}. We collected and labeled a corpus of 918 iOS app privacy policies, of which 67.3\% were legitimate privacy policies, and the rest were unrelated webpages. After training, we evaluated our classifier on an unseen test set with 64 positive examples and 42 negative examples. Our classifier was able to achieve 98.1\% accuracy with an F1 score of 98.4\% (precision = 100\%, recall = 96.9\%). To run our system efficiently, we only classified pages directly linked to by the provided privacy policy URL.

\subsection{Design of a Distributed Infrastructure}

\begin{figure}[h]
    \centering
    \includegraphics[width=0.7\linewidth]{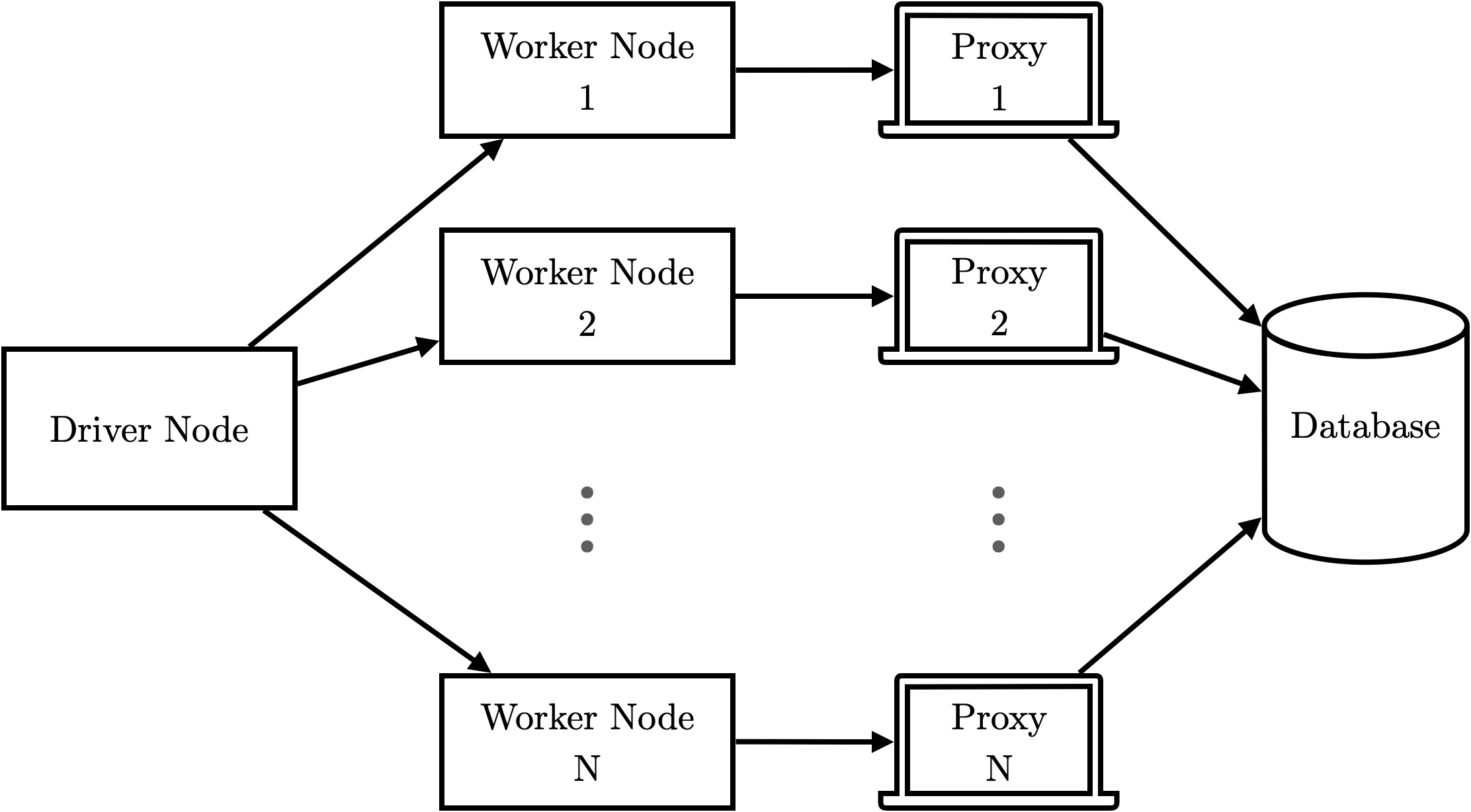}
    \caption[ATLAS Data Collection Pipeline]{A diagram of the ATLAS data collection pipeline.}
    \label{fig:atlas-infrastructure}
\end{figure}

The scale of our study necessitated the design, development, and deployment of a highly parallelizable and distributed data collection pipeline. This enabled us to collect all data within a relatively short window. To this end, we created an infrastructure as depicted in Figure \ref{fig:atlas-infrastructure}. We utilized a driver node to coordinate work between $N$ worker nodes. Each worker node ran a headless Firefox browser to replicate a real-world browser. This gave us the ability to capture dynamically loaded content and follow any webpage redirects -- emulating the experience of a real user. To avoid rate-limiting, we utilized a pool of $N$ proxy servers (one per worker) to increase the number of available IP addresses. The SOCKS5 protocol was used to connect worker nodes to proxies. After webpages were retrieved, they were saved in a shared database.

\begin{figure}[h]
    \centering
    \includegraphics[width=0.85\linewidth]{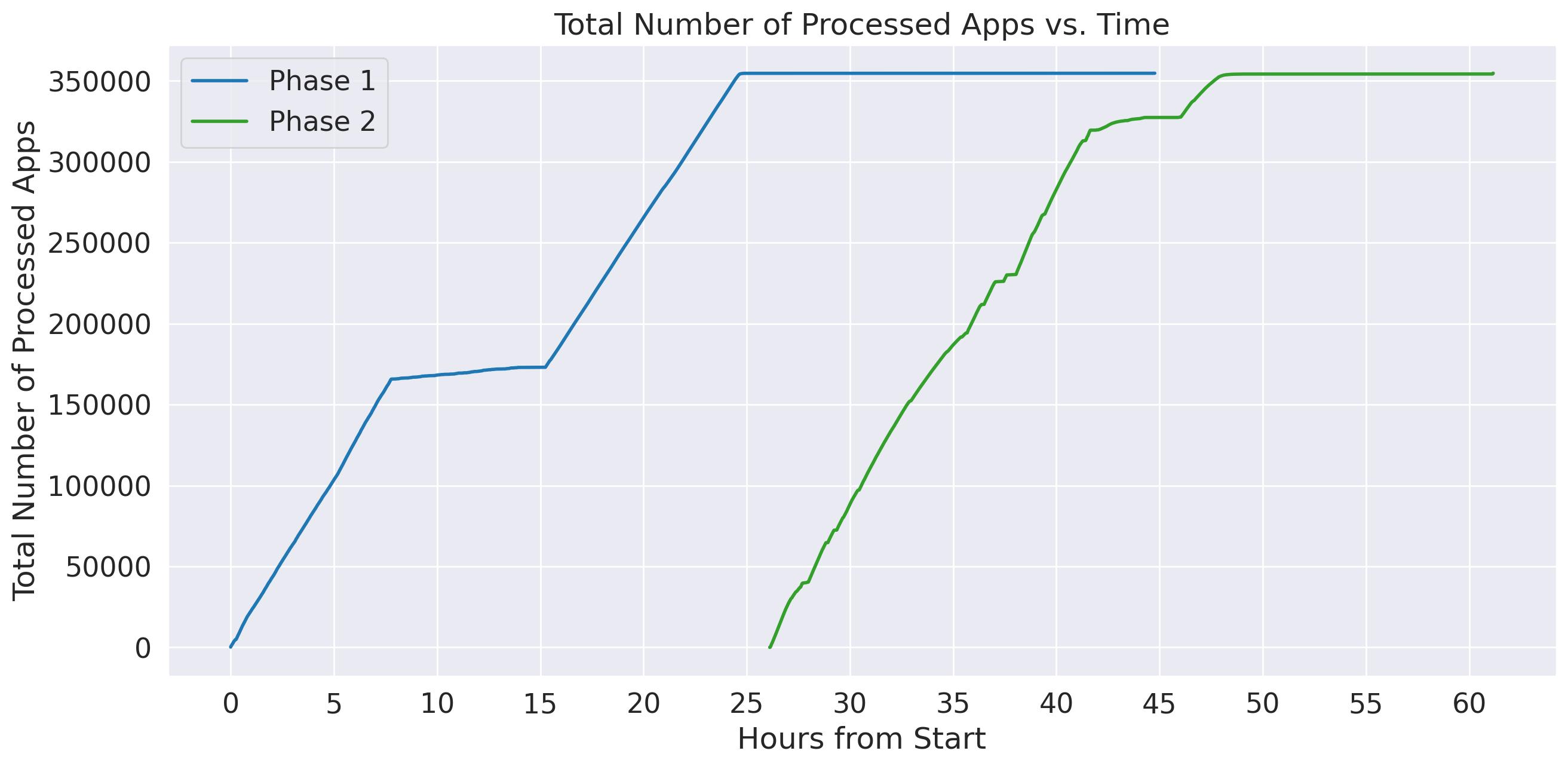}
    \caption[ATLAS Data Processing Rate]{A diagram of the number of processed apps over time. Data collection began on January 29$^{\text{th}}$, 2023 and ended on January 31$^{\text{st}}$, 2023.}
    \label{fig:atlas-processing-rate}
\end{figure}

We began data collection on January 29$^{\text{th}}$, 2023, and our system ran until January 31$^{\text{st}}$, 2023. We completed data collection in two phases. Phase 1 focused on downloading app listings from the iOS App Store, and Phase 2 focused on downloading privacy policies. We used slightly different configurations for each phase. Phase 1 utilized one driver node with 49 worker nodes and 49 proxies; whereas Phase 2 used one driver node with 80 worker nodes and no proxies. Proxies were not required in Phase 2 since privacy policies were hosted on different domains, so rate-limiting was not a concern.

As depicted in Figure \ref{fig:atlas-processing-rate}, Phase 1 ran at a rate of approximately 21,000 apps per hour. Around the 7.5-hour mark, our proxy servers initiated a nightly-reboot cycle, which caused the rate to diminish. After manual intervention, our system resumed around the 15-hour mark. 9 proxy servers were no longer responsive, so our rate slowed to approximately 19,000 apps per hour. Phase 2 began around hour 26 at an average rate of 20,500 apps per hour. At around hour 45, we reran Phases 1 and 2 in an attempt to re-download app listings and privacy policies that timed out. We reran Phase 2 once more around the 60-hour mark to collect the final set of privacy policies. In total, we were able to collect 345,725 apps.

\subsection{Privacy Policy Accessibility and Privacy Label Adoption}

\begin{figure}[h!]
    \centering
    \includegraphics[height=1.1in]{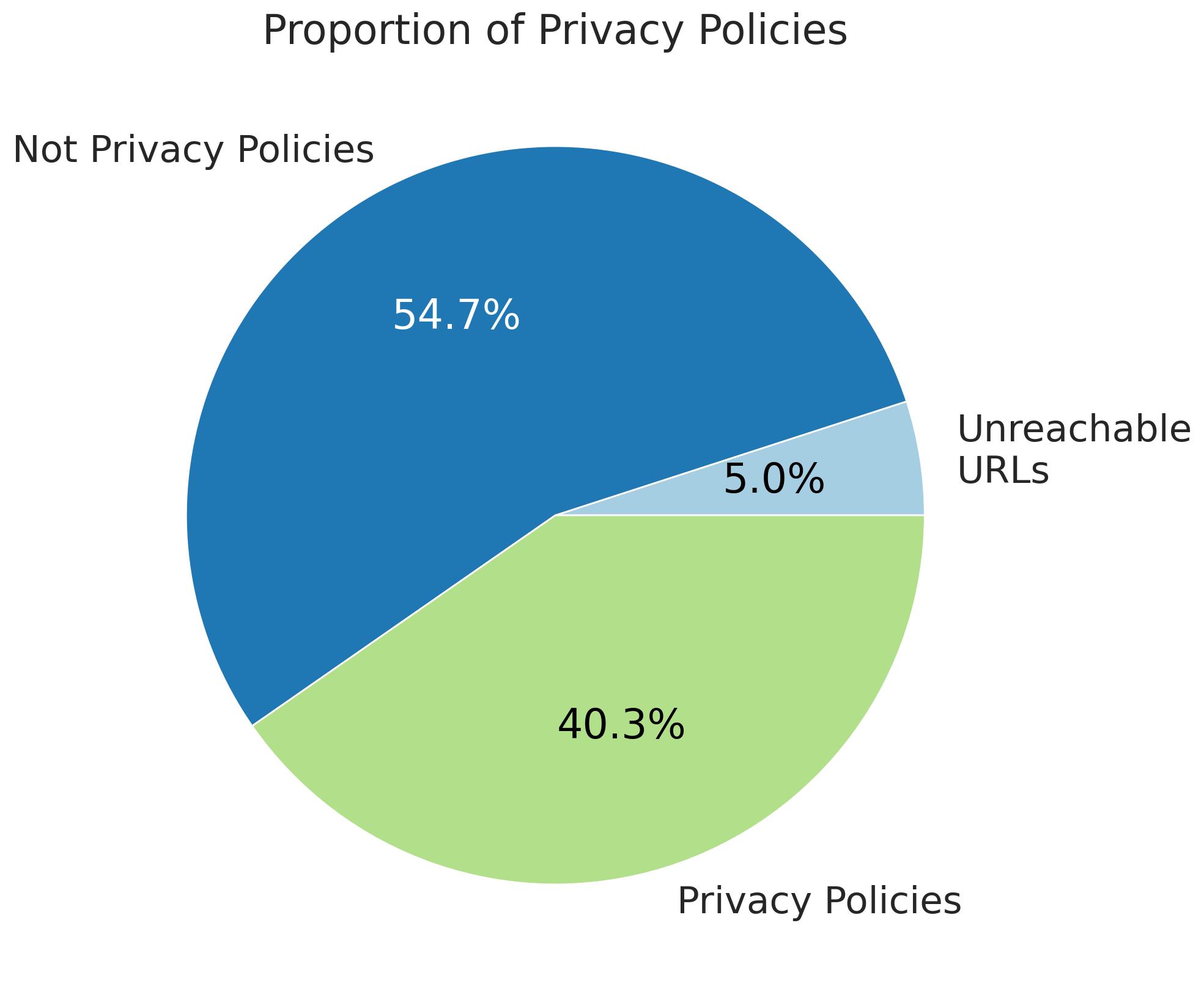}
    \includegraphics[height=1.1in]{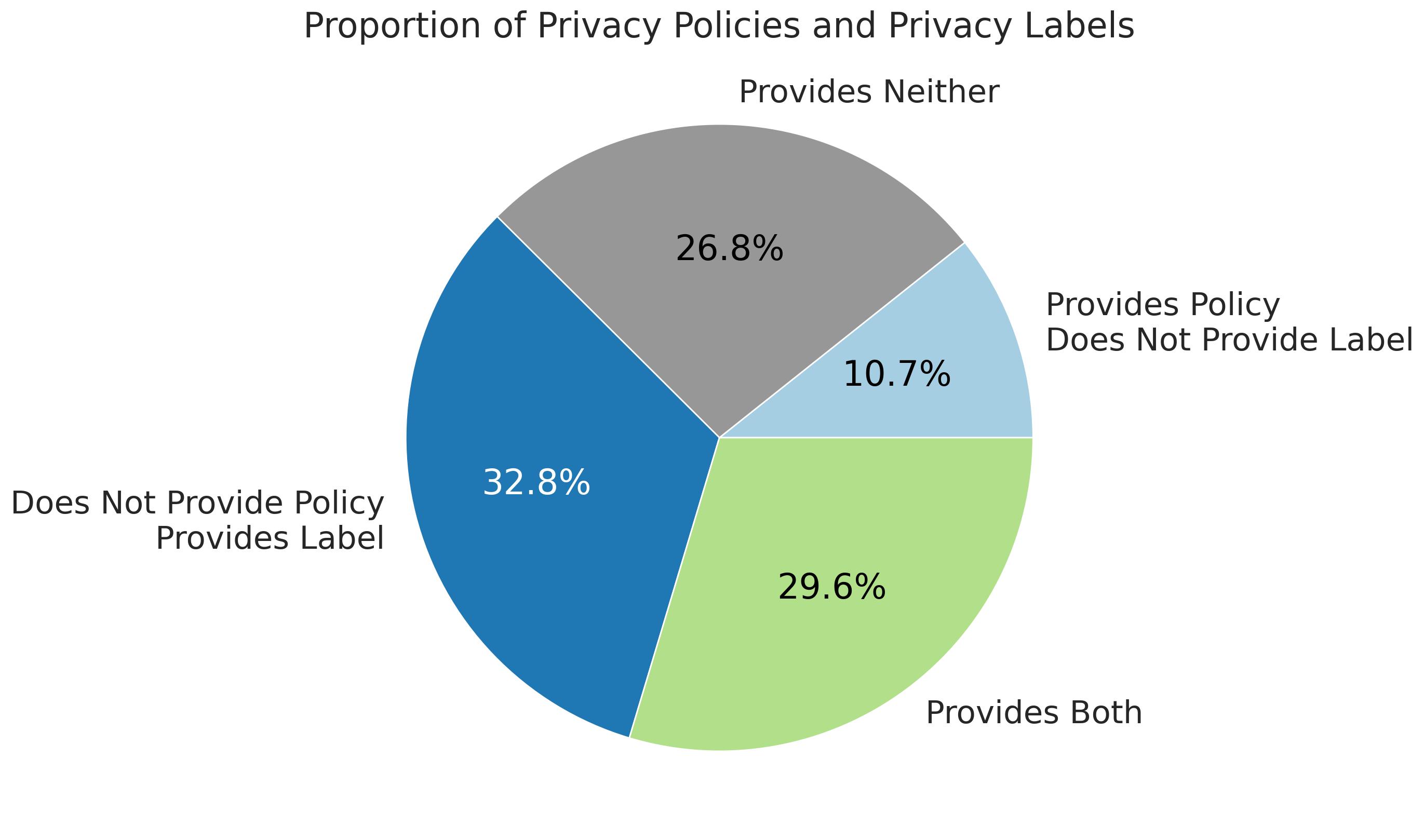}

    \caption[Proportion of Privacy Policies and Privacy Labels in the iOS App Store]{The chart on the left depicts the proportion of webpages linked to by reported privacy policy URLs in the iOS App Store. The chart on the right depicts the proportion of apps providing direct links to privacy policies and privacy labels in the iOS App Store.}
    \label{fig:privacy-policy-label-distribution}
\end{figure}

We found several interesting trends upon analysis of downloaded apps. Even though apps are required to link to privacy policies, a substantial number of apps provided extraneous links instead. As depicted in Figure \ref{fig:privacy-policy-label-distribution}, 5.0\% of apps provided dead links. Moreover, 54.7\% of apps provided links that led to extraneous webpages, such as landing pages, home pages, and 404s. Only 40.3\% of apps provided direct links to legitimate privacy policies, which we characterized as accessible privacy policies.

Next, we analyzed the adoption rate of privacy labels. Apple began requiring apps published or updated after December 2020 include privacy labels. Interestingly, we discovered that 62.5\% of sampled apps provided privacy labels. A number substantially higher than those that provided accessible privacy policies. As depicted in Figure \ref{fig:privacy-policy-label-distribution}, only 29.6\% of apps (105,131 apps) provided both accessible privacy policies and privacy labels. We focused on this subset of apps to conduct our compliance analysis.

Finally, we analyzed the most common types of reported data collection, as show in Figure \ref{fig:apps-per-data-type}. Unsurprisingly, the most common data type collected was Crash Data, followed by Product Interaction and Email Address. Interestingly, Gameplay Content was the second least reported type of data collected, even though Games were the most common type of app in our dataset (and in the App Store). Also indicated is the number of popular apps reporting collection of each data type: the distribution of collection across data types for popular apps approximately follows the same distribution as apps available in the broader App Store.
\section{Autogenerating Privacy Labels from Privacy Policies}
\label{chap:4-generating}

We next focused our efforts on predicting privacy labels from privacy policies. We formulated the task as a supervised multi-class, multi-label document classification problem, where the input was a privacy policy, and the output was a set of collected data types, as indicated by the privacy policy. We targeted the same 32 data types included in iOS privacy labels. To reduce complexity, we only identified which data types were collected, and not how they were used. We also treated predicting each data type as an independent binary classification problem: we created 32 models, where each was responsible for predicting a single data type.

\subsection{Dataset Construction}

Unlike prior research, which used high quality annotator labeled privacy policies, we relied solely on developer reported privacy labels to learn from privacy policies \cite{story2019natural} \cite{zimmeck2019maps}. We began by filtering our downloaded set of apps to those which provided both privacy policies and privacy labels. This left us with 105,131 iOS apps. However, upon analysis, we found many privacy policies to be shared among apps -- developers likely reused the same privacy policy among many of their apps. We determined policy uniqueness by comparing privacy policy URLs for exact matches. %
In total, we found 34.5\% of policies to be duplicates, leaving us with 68,863 unique privacy policies.

Care was taken to preserve the structure of differing privacy labels for identical privacy policies. We assumed that shared privacy policies are written generally, so as to be applicable to multiple apps. Privacy labels, however, are constructed on a per app basis; therefore, they only represent a subset of the privacy policy. As a result, we reasoned that when duplicate policies were merged together, their privacy labels needed to be combined by taking the union of collected data types. %

\subsection{Sampling Procedure}\label{sec:sampling_procedure}

While using developer reported privacy labels enabled us to build a training corpus almost a hundred times larger than previous work, our data was expected to be somewhat ``noisy'' \cite{story2019natural} \cite{zimmeck2019maps}. %
Smaller scale analyses in the past suggest that privacy labels may not be accurate \cite{li2022understanding} \cite{gardner2022helping}. While much work has been dedicated to learning from noisy, mislabeled data (\cite{song2022learning}, \cite{zhao2021evaluating}), it is ``unfair and unreasonable to have noise in the'' testing data \cite{blaheta2002handling}. We outline a technique for importance sampling to help reduce noise when creating training and testing datasets. For clarity, we define a \textbf{Positive} instance as a privacy policy associated with a privacy label that reported a particular data type as being collected. A \textbf{Negative} instance is where the data type is reported to not be collected.

\begin{figure}[h]
    \begin{subfigure}[t]{0.32\linewidth}
        \centering
        \includegraphics[width=\linewidth]{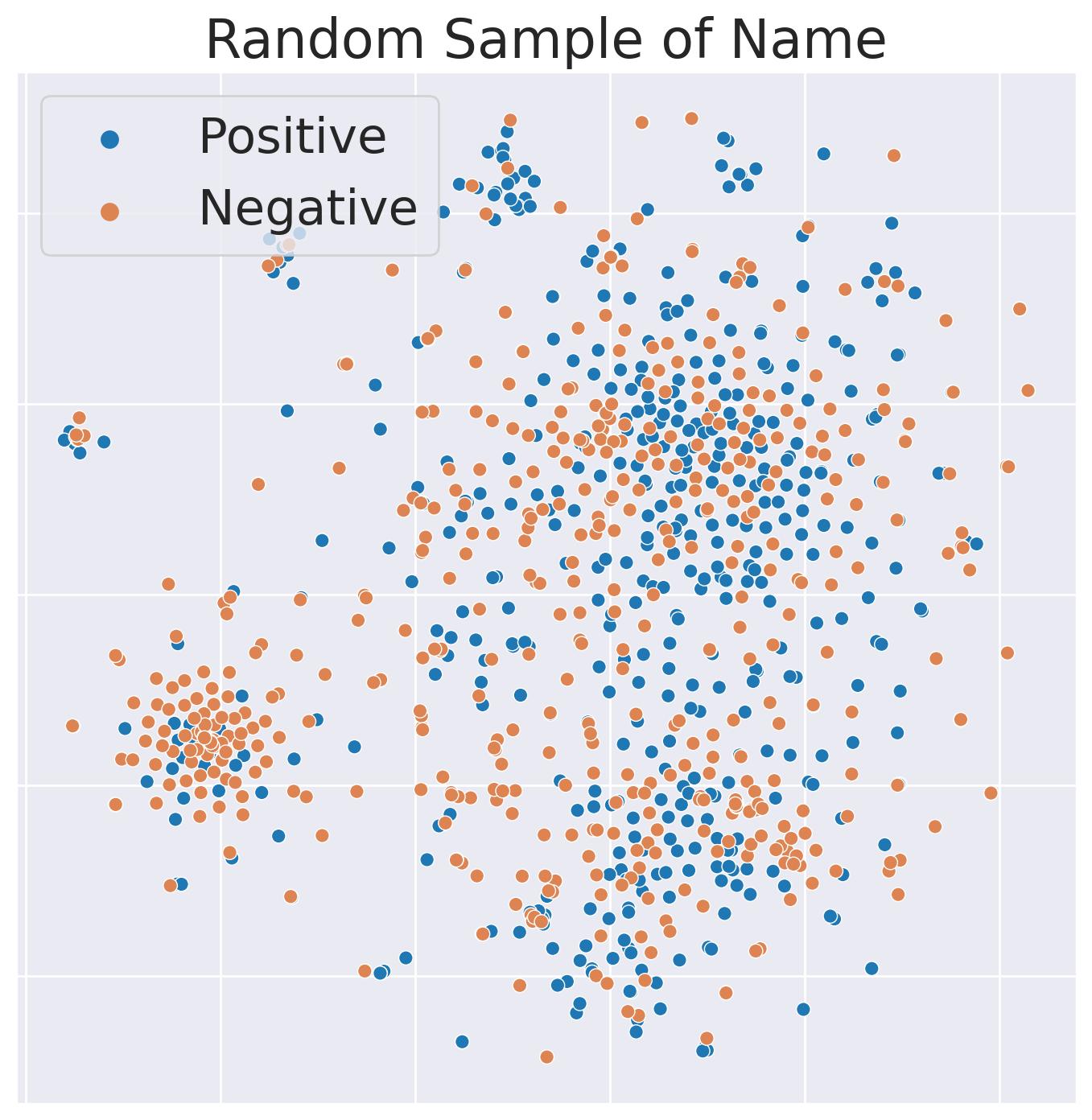}
        \caption{Random Sample}
        \label{fig:random-sampling}
    \end{subfigure}
    \begin{subfigure}[t]{0.32\linewidth}
        \centering
        \includegraphics[width=\linewidth]{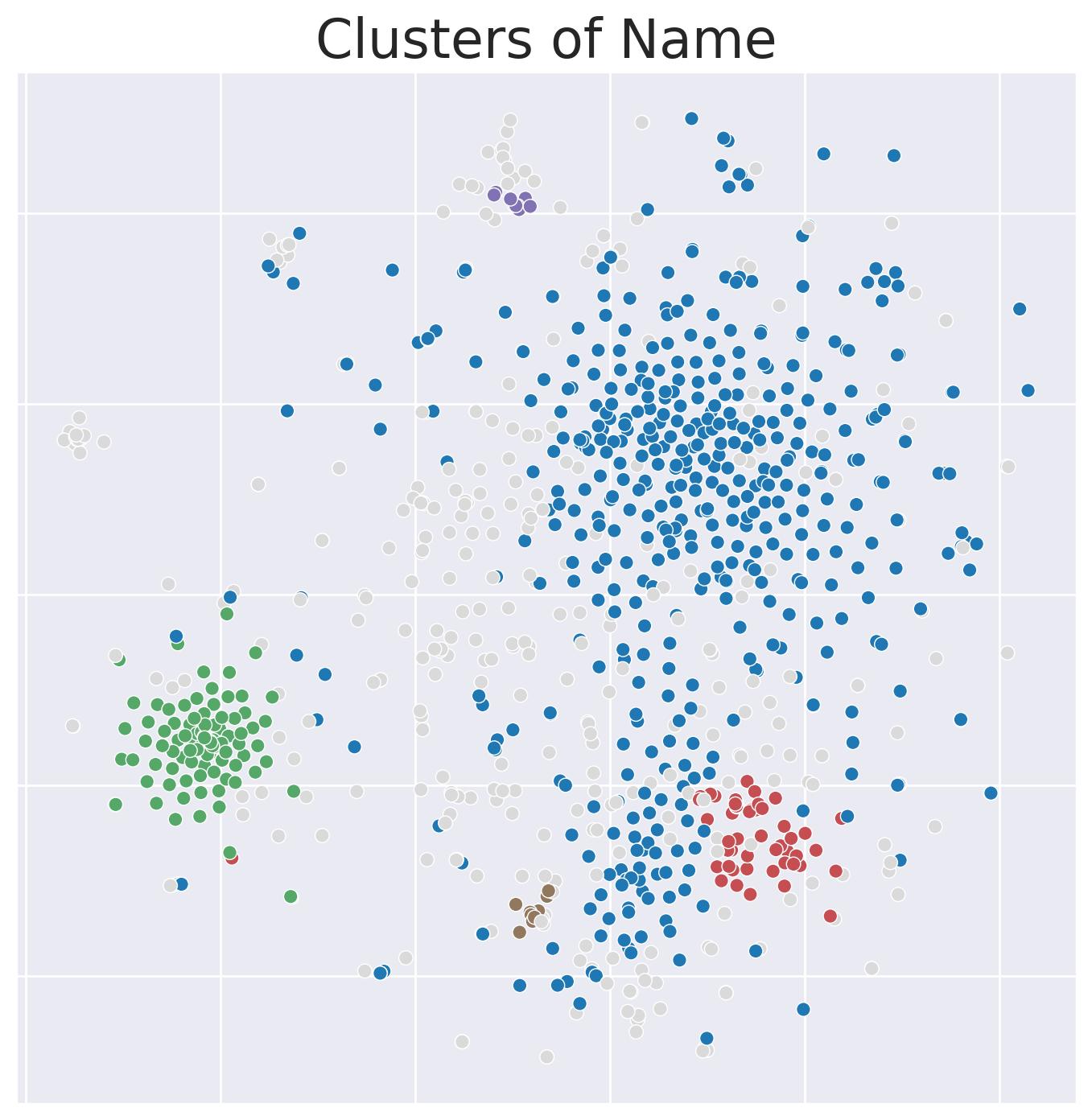}
        \caption{Clusters}
        \label{fig:random-sampling-clusters}
    \end{subfigure}
    \begin{subfigure}[t]{0.32\linewidth}
        \centering
        \includegraphics[width=\linewidth]{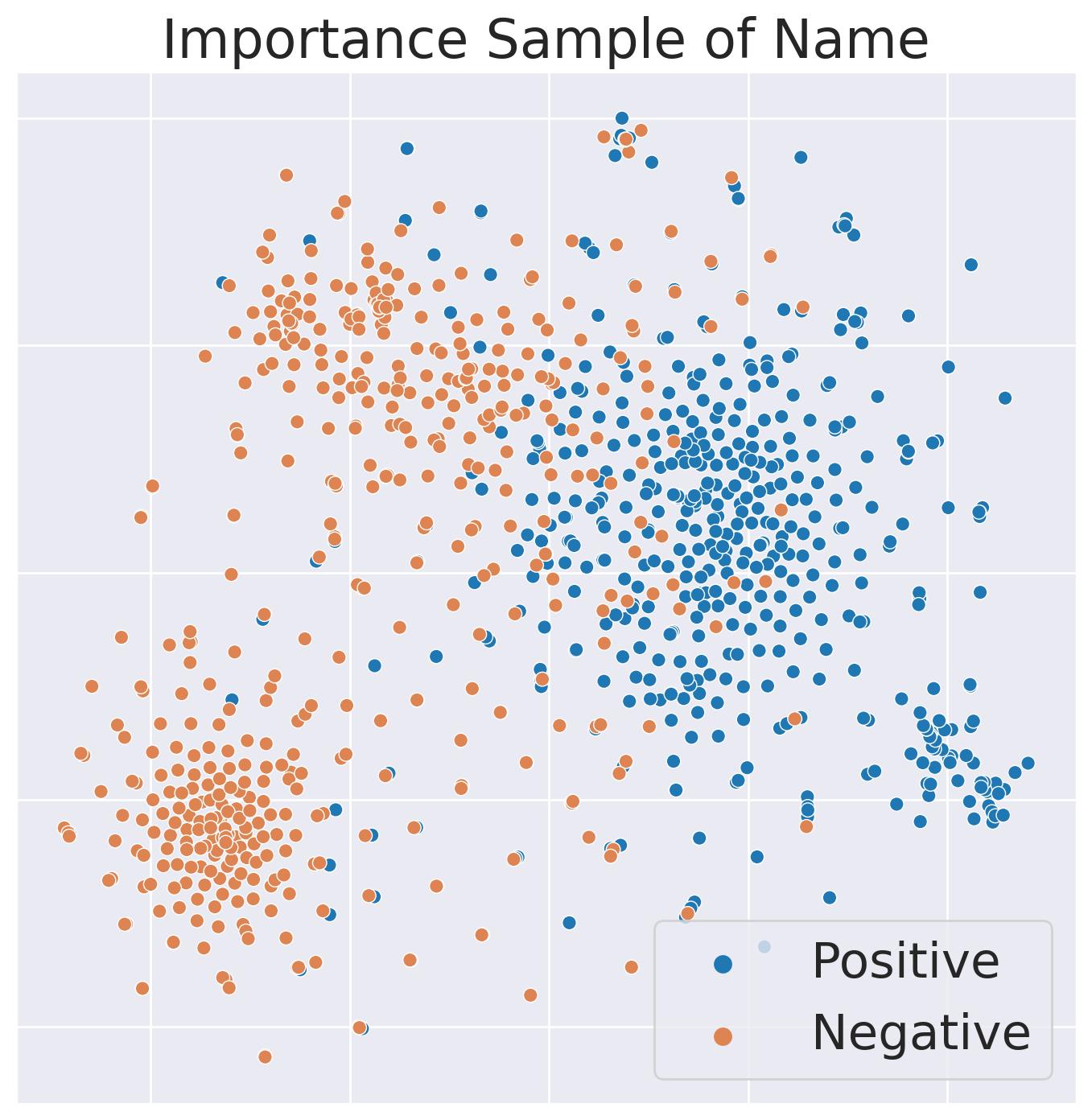}
        \caption{Importance Sample}
        \label{fig:importance-sampling}
    \end{subfigure}
    \caption[Random Sampling]{On the left is a a random sample of privacy policies declaring / not declaring collection of the Name data type.  %
    In the center is a depiction of clusters appearing within the random sample. On the right is an importance sample of the Name data type. For all three plots, each point is a t-SNE representation of a privacy policy TF-IDF embedding.
    }
\end{figure}

Figure \ref{fig:random-sampling} characterizes how noisy the underlying data was by providing a depiction of a random sample of the Name data type: we randomly sampled 500 policies whose privacy label declared collection of Name, and 500 policies that did not. We then extracted an embedding from each policy using TF-IDF vectorization and created a two-dimensional representation using t-SNE \cite{van2008visualizing}. There is significant overlap between positive and negative instances and no clear distinction between the two groups; this suggests that many policies might be mislabeled.

However, as depicted in Figure \ref{fig:random-sampling-clusters}, natural clusters tended to appear within the randomly sampled data. Points were clustered using DBSCAN -- since the number of clusters was not known a priori and the underlying data contains some noise -- after performing Latent Semantic Analysis (LSA) on the high dimensional TF-IDF embeddings \cite{ester1996density} \cite{dumais2004latent}. LSA was performed by using TruncatedSVD to project the high-dimensional data onto 10 components. We reasoned that these clusters are semantically similar privacy policies and should therefore all have the same privacy label (either positive or negative). For example, if a cluster had 20 positive instances and 5 negative instances, then the negative instances could potentially be mislabeled, since the content of their privacy policies were similar to many positive instances.

We assigned clusters a positive or negative label by conducting a two-proportion z-test that compared the incidence of positive to negative labels within a cluster. If a cluster had a statistically significantly larger proportion (i.e., p $<$ 0.05) of one class than the other, we assigned the cluster the label of the more prominent class. In cases where determining cluster labels was inconclusive (i.e., p $\geq$ 0.05), we disregarded the cluster entirely. We then computed centroids for all positive and negative clusters.

Finally, to construct a dataset of size $N$, with $\frac{N}{2}$ positive and $\frac{N}{2}$ negative examples, we conducted a random oversample of $N$ instances per class. Then, for each class, we selected the $\frac{N}{2}$ examples closest to the class centroids, effectively sampling points closer to centroids with higher probability than those farther away. We also equally weighted the contribution of each class centroid. Each of the $C$ centroids had approximately $\frac{N}{2C}$ examples associated with it. To account for edge cases where no clusters were identified, we fell back to a simple random sample.

This process was used to construct the test ($N=150$), validation ($N=150$), and training sets ($N=1000$), for each data type. We had 4,376 unique examples within the test sets, 4,262 unique examples within the validation sets, and 22,262 unique examples within the training sets. In total, we used 30,900 unique privacy policies to train and evaluate our classifiers.

Intuitively, our sampling procedure can be thought of as focusing on high-density areas with large amounts of information more likely to be labeled correctly. Points on the edge have a lower probability of being selected: they are assumed to be less likely to be labeled correctly, therefore contributing less useful information. Figure \ref{fig:importance-sampling} demonstrates how importance sampling creates a clear separation between the positive and negative classes, in contrast to random sampling shown in Figure \ref{fig:random-sampling}.

\subsection{Model Selection}

After constructing the datasets, we trained using several model architectures. We used logistic regression as our baseline architecture, with a similar configuration to prior work \cite{story2019natural}. We then graduated to using more complex architectures: multilayer perceptron (MLP), RegLSTM, BERT, RoBERTa, and Longformer. We conducted extensive hyperparameter tuning for each model, per data type. Since we were training a separate model per data type, we created a final ``ensemble'' model utilizing a combination of architectures that maximized validation Macro F1 Score.

Models were trained using PyTorch on four NVIDIA GeForce RTX 2080 Ti GPUs \cite{paszke2017automatic}. In total, it took approximately 48 hours to hyperparameter tune and train 32 models across 6 different architectures.

\begin{figure}[h]
    \centering
    \includegraphics[height=1.5in]{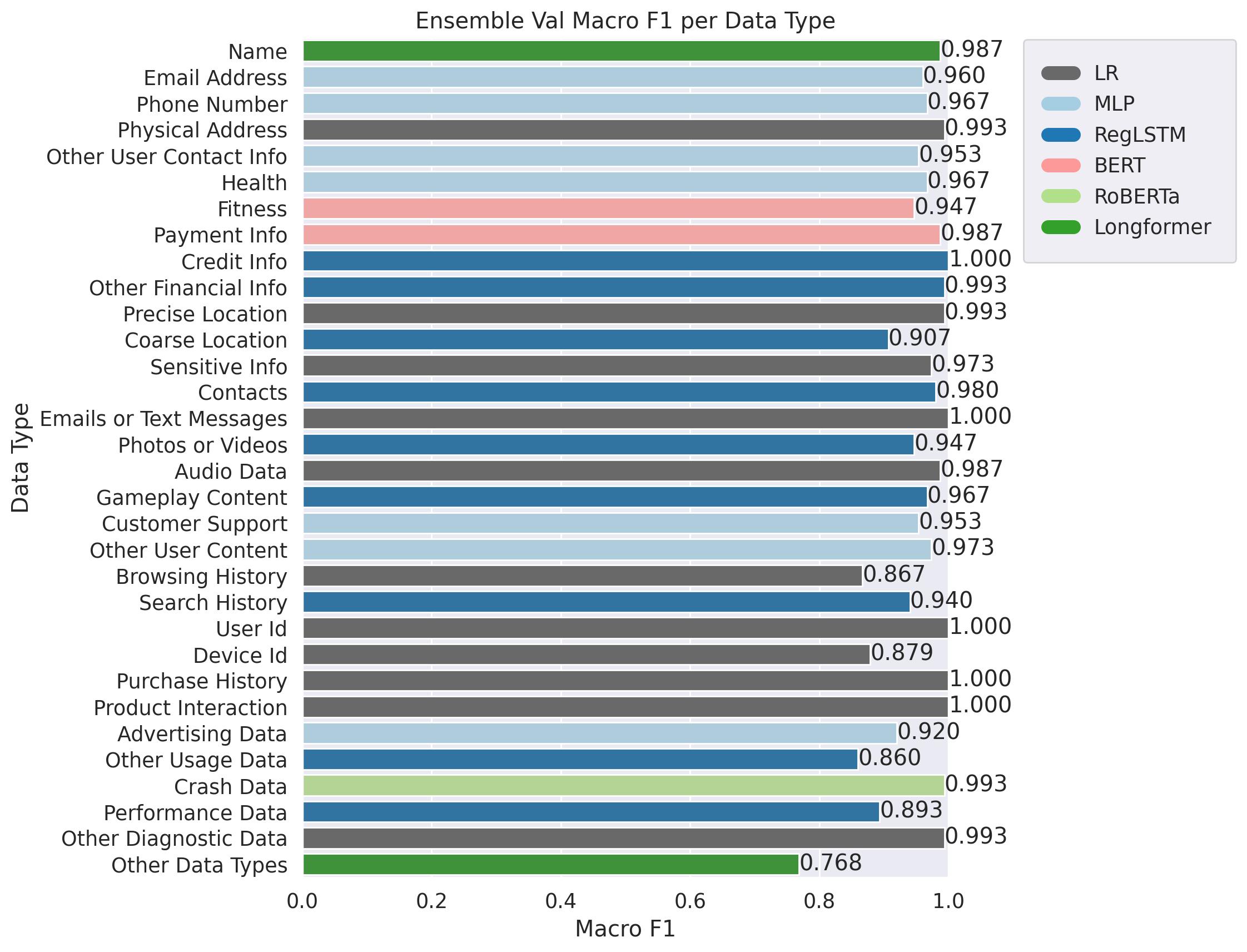}
    \includegraphics[height=1.5in]{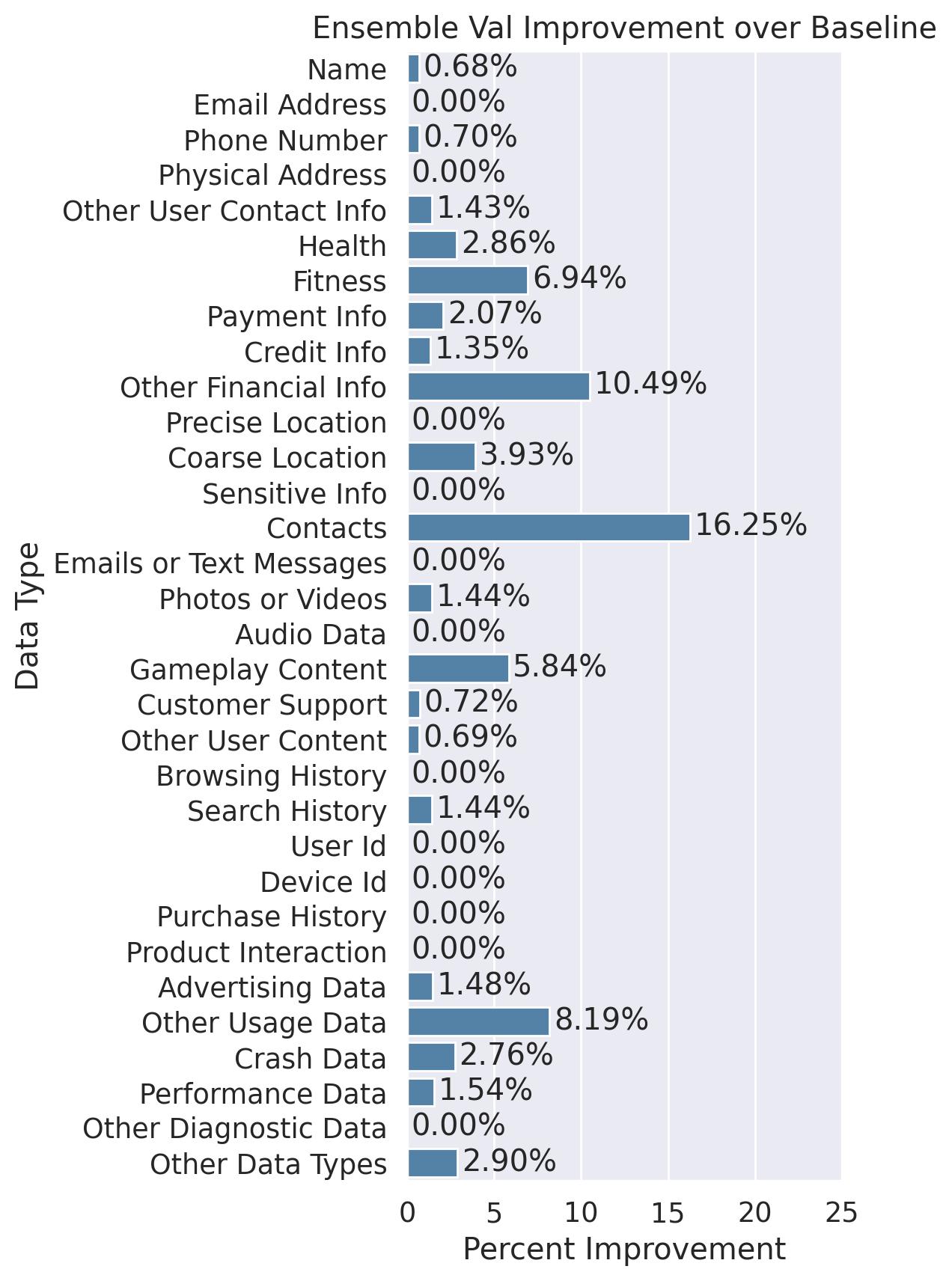}
    \caption[Ensemble Validation Performance]{Ensemble validation results per data type are shown on the left, and improvement over the baseline model is shown on the right.}
    \label{fig:ensemble-val-perf}
\end{figure}

Our final model was constructed with an ensemble of architectures. Since each data type was trained separately, we selected the architecture that maximized the Macro F1 score for that data type. In the case that multiple models have identical Macro F1 scores, we selected the simplest model. Figure \ref{fig:ensemble-val-perf} shows the selection of architecture per data type, as well as overall improvement over the baseline architecture. No singular architecture dominated; however, the baseline architecture was highly competitive compared to more complex architectures.

\subsection{Model Performance}

\begin{figure}[h]
    \centering
    \includegraphics[height=2.4in]{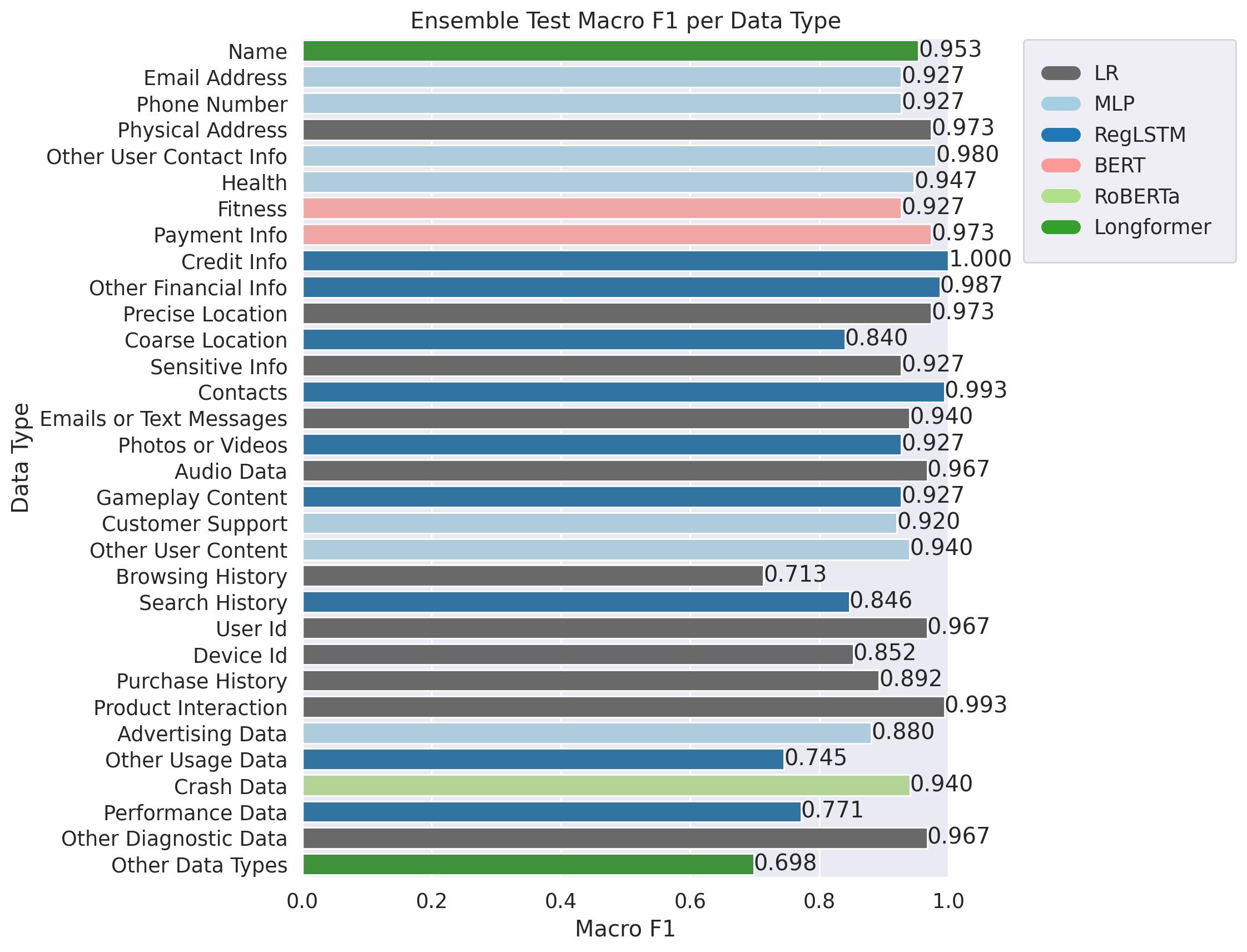}
    \caption[Ensemble Test Performance]{The final ensemble-based model test results per data type.}
    \label{fig:ensemble-test-perf}
\end{figure}

Finally, we evaluated our ensemble of models on an unseen set of test data. Figure \ref{fig:ensemble-test-perf} shows the test Macro F1 score achieved per data type. Overall, we were able to achieve an average accuracy of 91.3\% and an average Macro F1 score of 91.3\% across all classes. Notably, we were able to achieve a Macro F1 score of 100\% for Credit Info, with several other data types being the high-90s. Table \ref{tab:ensemble-test-perf} provides detailed statistics about model performance on the test dataset.

\section{Discrepancy Analysis}
\label{chap:5-compliance}

After training our ensemble-based classifier, we used it to predict privacy labels for the remaining privacy policies. After removing training data, we were left with a set of privacy policies corresponding to 61,596 iOS apps. The following analysis is presented for those apps.

\subsection{Characterizing Potential Compliance Issues}

We first begin by characterizing a potential compliance issue. For an arbitrary app, let $P$ be the set of data types collected by the app as disclosed in its privacy policy, and let $L$ be the set of data types collected by the app as disclosed in its privacy label. Let $D$ represent the set of 32 data types captured by iOS privacy labels, and let $d \in D$ be a single data type. We use $\hat{d}$ to denote a predicted data type. By definition, we can write $P \subseteq D$ and $L \subseteq D$.

The first type of potential compliance issue is an \textbf{Incomplete Privacy Policy} (often shortened to Incomplete Policy in the rest of the section): $$\exists \, d \in D \text{, such that } (\hat{d} \not \in P) \land (d \in L)$$ Intuitively, an incomplete policy means that a privacy policy does not disclose the collection of a data type, while its developer reported privacy label does.

The second type of potential compliance issue is an \textbf{Incomplete Privacy Label} (often shortened to Incomplete Label): $$\exists \, d \in D \text{, such that }  (\hat{d} \in P) \land (d \not \in L)$$ Intuitively, an incomplete label is when collection of a data type is disclosed within a privacy policy, but not within the developer reported privacy label.

Potential compliance issues are on a per data type basis. This means that the total number of potential compliance issues per app in our analysis is capped at 32 (one per data type).

We also took care to ensure that we only identified potential compliance issues with high probability. Figure \ref{fig:pdfs-name-precise-location} shows the probability that a policy collects a particular data type (i.e. $p_{\hat{d} \in P}$). Since we formulated this as a binary classification problem for each data type, $p_{\hat{d} \not \in P} = 1 - p_{\hat{d} \in P}$. For example, as shown in \ref{fig:pdf-name}, most policies have a high probability of collecting Name or a high probability of not collecting Name; however, for many policies, it is uncertain that they collect Precise Location (Figure \ref{fig:pdf-precise-location}), as many probabilities are near 50\%. So, we only considered
\begin{align*}
    (\hat{d} \not \in P) &\Longleftrightarrow (p_{\hat{d} \in P} < 0.25)\\
    (\hat{d} \in P) &\Longleftrightarrow (p_{\hat{d} \in P} > 0.75)\\
\end{align*}
which is depicted by the two vertical lines in Figures \ref{fig:pdf-name} and \ref{fig:pdf-precise-location}. Intuitively, we only considered predictions with high probability; otherwise, we classify the prediction as ``Inconclusive.''

\subsection{Potential Compliance Issues by Data Type}

\begin{figure*}[h]
    \centering
    \begin{subfigure}{0.49\linewidth}
        \centering
        \includegraphics[height=2.75in]{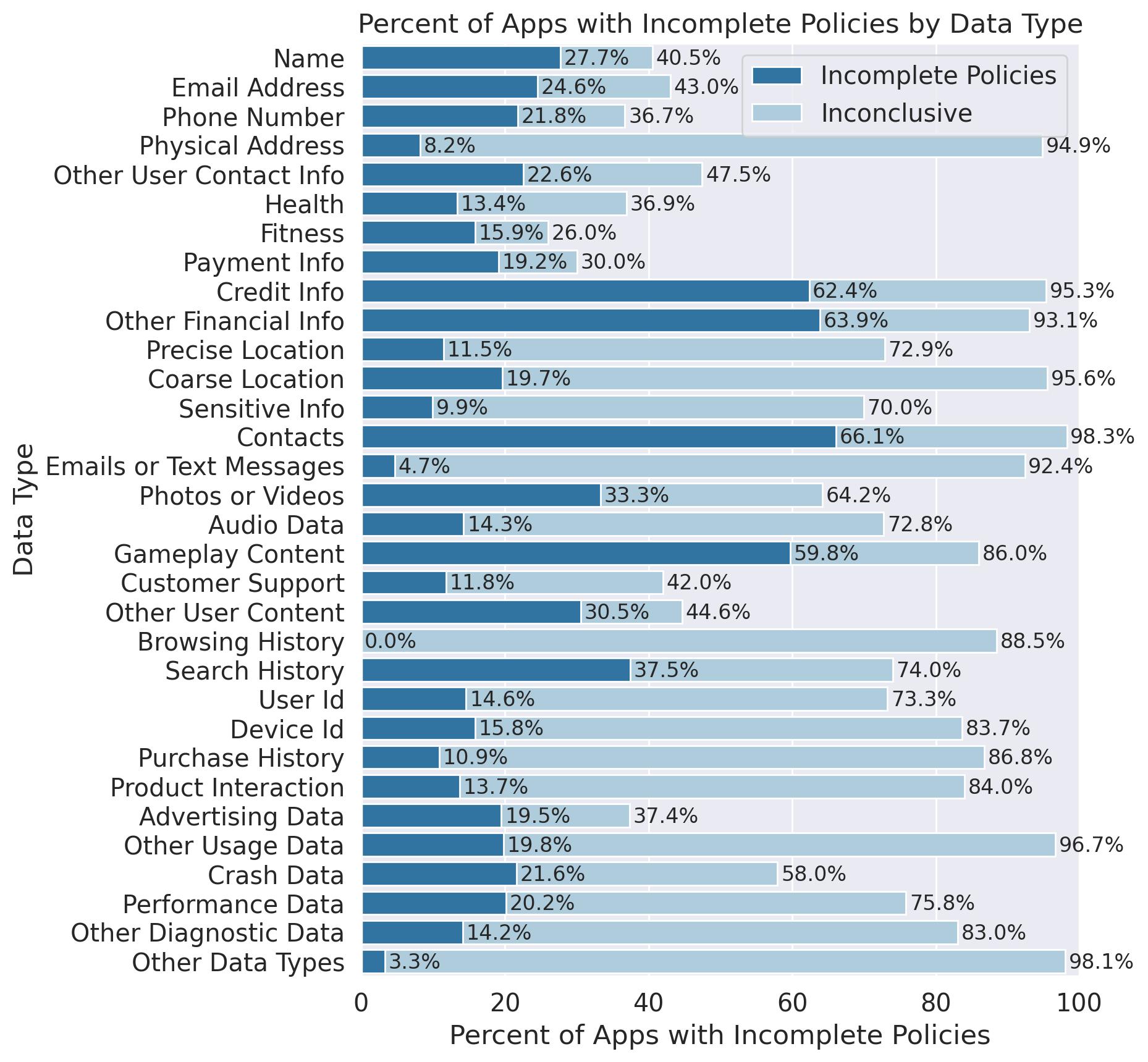}
        \caption{Incomplete Policies}
        \label{fig:incomplete-policies-data-type}
    \end{subfigure}
    \begin{subfigure}{0.49\linewidth}
        \centering
        \includegraphics[height=2.75in]{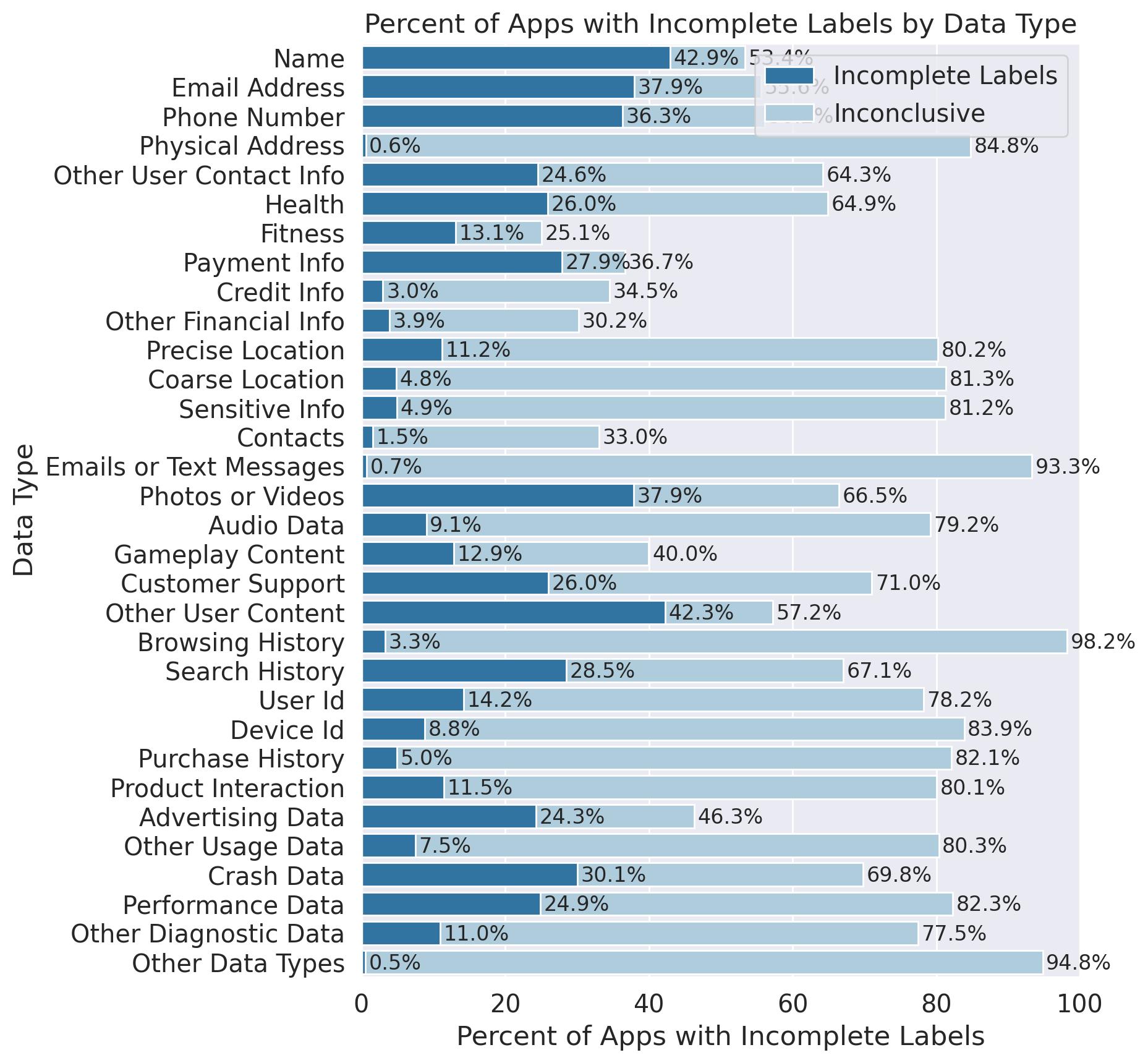}
        \caption{Incomplete Labels}
        \label{fig:incomplete-labels-data-type}
    \end{subfigure}
    \caption[Potential Compliance Issues by Data Type]{Potential compliance issues by data type. The bar for each data type is stacked with two segments. The first segment (in dark blue) indicates potential compliance issues (i.e. high probability predictions). The second segment (in light blue) indicates inconclusive results (i.e. low probability predictions).}
    \label{fig:error-rates-data-type}
\end{figure*}

We offer a breakdown of compliance issues by data type, as show in Figure \ref{fig:error-rates-data-type}. Figure \ref{fig:incomplete-policies-data-type}, shows the rate of incomplete policies by data type. The percentage for each data type, $d \in D$, is the total number of incomplete policies for $d$ divided by the total number of apps where $d \in L$. For example, 27.7\% of apps have privacy policies that do not declare collection of Name, even when their privacy labels do.

Similarly, Figure \ref{fig:incomplete-labels-data-type} shows the rate of incomplete labels by data type. The percentage for each data type, $d \in D$, is the total number of incomplete labels for $d$ divided by the total number of apps where $d \not \in L$. For example, 42.9\% of apps have privacy labels that do not declare collection of Name, even when their privacy policies do.

Of note are the high rates of incomplete policies within financial disclosures: 62.4\% of apps declaring Credit Info and 63.9\% of apps declaring Other Financial Info on their privacy labels have incomplete privacy policies. This particular set of potential compliance issues could be a violation of the Gramm-Leach-Bliley Act \cite{ftc2023gramm}.

\subsection{Distribution of Potential Compliance Issues}\label{sec:dist-compliance-issues}

\begin{table}[h]
    \centering
    \caption[Cumulative Distribution of Potential Compliance Issues]{This table provides a cumulative distribution of potential compliance issues. \label{tab:cdf-errors}}
    \footnotesize
    \begin{tabular}{p{0.15\linewidth}p{0.20\linewidth}p{0.20\linewidth}p{0.20\linewidth}}
        \toprule
        {} & \textbf{Incomplete Policies (\%)} & \textbf{Incomplete Labels (\%)} & \textbf{Both (\%)} \\
        \toprule
        \textbf{1 or more}  &  26.8  &  85.7  & 88.0 \\
        \textbf{2 or more}  &  14.3  &  72.9  & 76.1 \\
        \textbf{3 or more}  &  8.5   &  61.3  & 65.0 \\
        \bottomrule
    \end{tabular}
\end{table}

Figure \ref{fig:distribution-cdf-compliance-issues} shows the distribution and cumulative distribution of potential compliance issues. In particular, incomplete policies are far less common than incomplete labels, with apps having an average of 0.62 incomplete policy discrepancies and 4.70 incomplete label discrepancies. When looking at the combination of incomplete policy and label errors, apps have 5.32 potential compliance issues on average. %

Figure \ref{fig:cdf-compliance-issues} depicts the CDF for each data type. Of note, 26.8\% of apps have at least one incomplete policy discrepancy and 85.7\% of apps have at least one incomplete label discrepancy. When analyzing both incomplete policies and labels, 88.0\% of apps have at least one discrepancy. We provide additional percentages in Table \ref{tab:cdf-errors}.

\subsection{App Rating vs Number of Potential Compliance Issues}

\begin{figure}[h]
    \centering
    \begin{subfigure}{0.49\linewidth}
        \centering
        \includegraphics[height=1.85in]{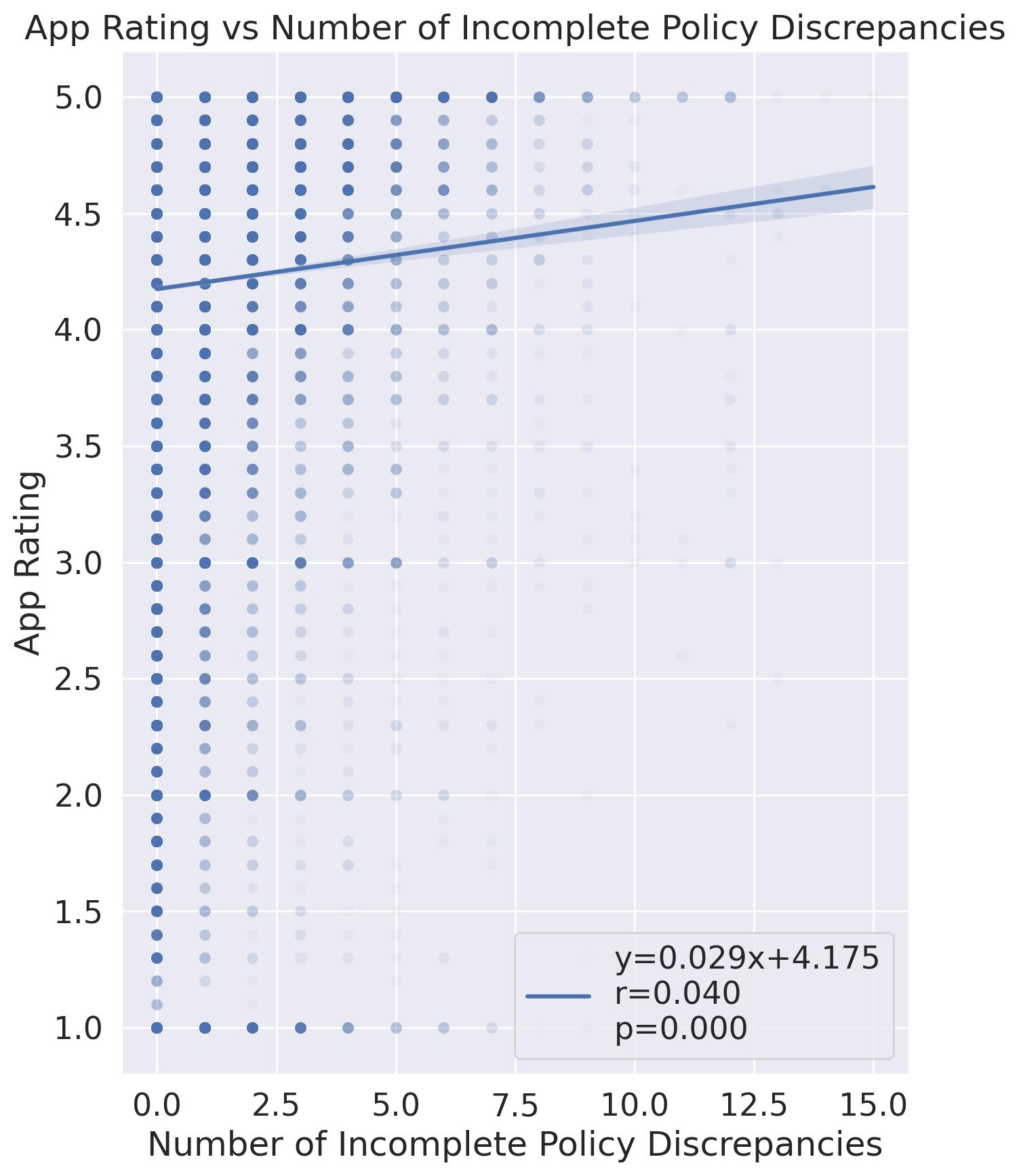}
        \caption{Rating vs Incomplete Policy Discrepancies}
        \label{fig:regression-policy-errors}
    \end{subfigure}
    \begin{subfigure}{0.49\linewidth}
        \centering
        \includegraphics[height=1.85in]{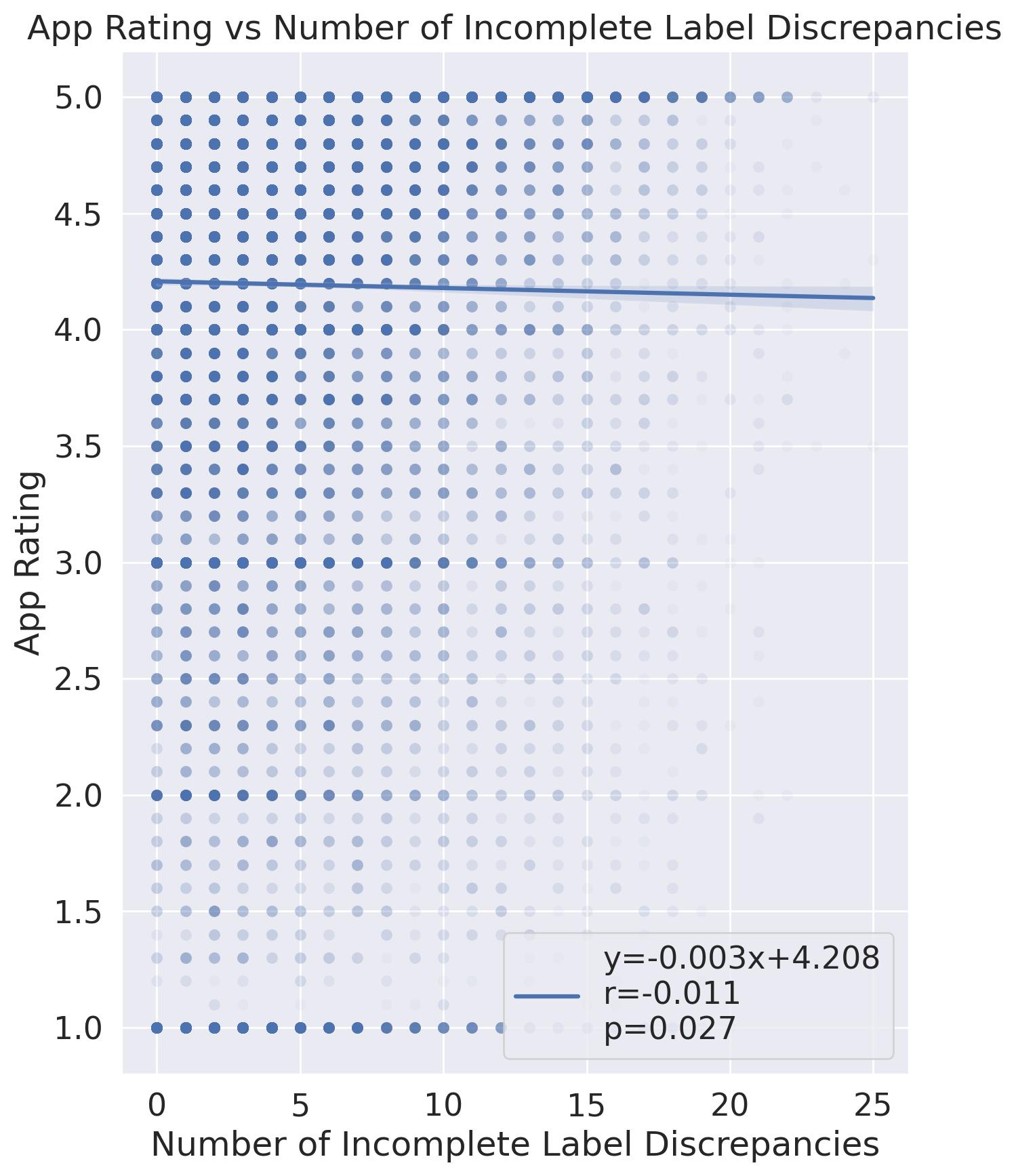}
        \caption{Rating vs Incomplete Label Discrepancies}
        \label{fig:regression-label-errors}
    \end{subfigure}
    \caption[App Rating vs Number of Compliance Issues]{The graph on the left depicts the correlation with respect to incomplete policy discrepancies, and the graph on the right depicts the correlation with respect to incomplete label discrepancies. Darker points represent a larger number of apps (iOS App Store ratings are discretized in tenths and discrepancies are integers). The shaded region around each line represents a 95\% confidence interval.}
    \label{fig:regression-errors}
\end{figure}

We also compared the trend between app rating and number of potential compliance issues, as shown in Figure \ref{fig:regression-errors}. Surprisingly, for incomplete policy compliance issues, we found a weak, significant \textit{positive} correlation ($r=0.04$, $p < 0.05$) between app rating and number of incomplete policy discrepancies. That is, as the number of incomplete policy discrepancies increase, the app rating tends to \textit{increase}.

While the previous result was surprising, the relationship between app rating and incomplete label discrepancies is as expected: we found a weak, significant negative correlation ($r = -0.011$, $p < 0.05$). That is, as the number of incomplete label discrepancies increase, the app rating tends to decrease.

\subsection{Potential Compliance Issues Among Popular vs Other Apps}

\begin{figure}[h]
    \centering
    \includegraphics[height=1.75in]{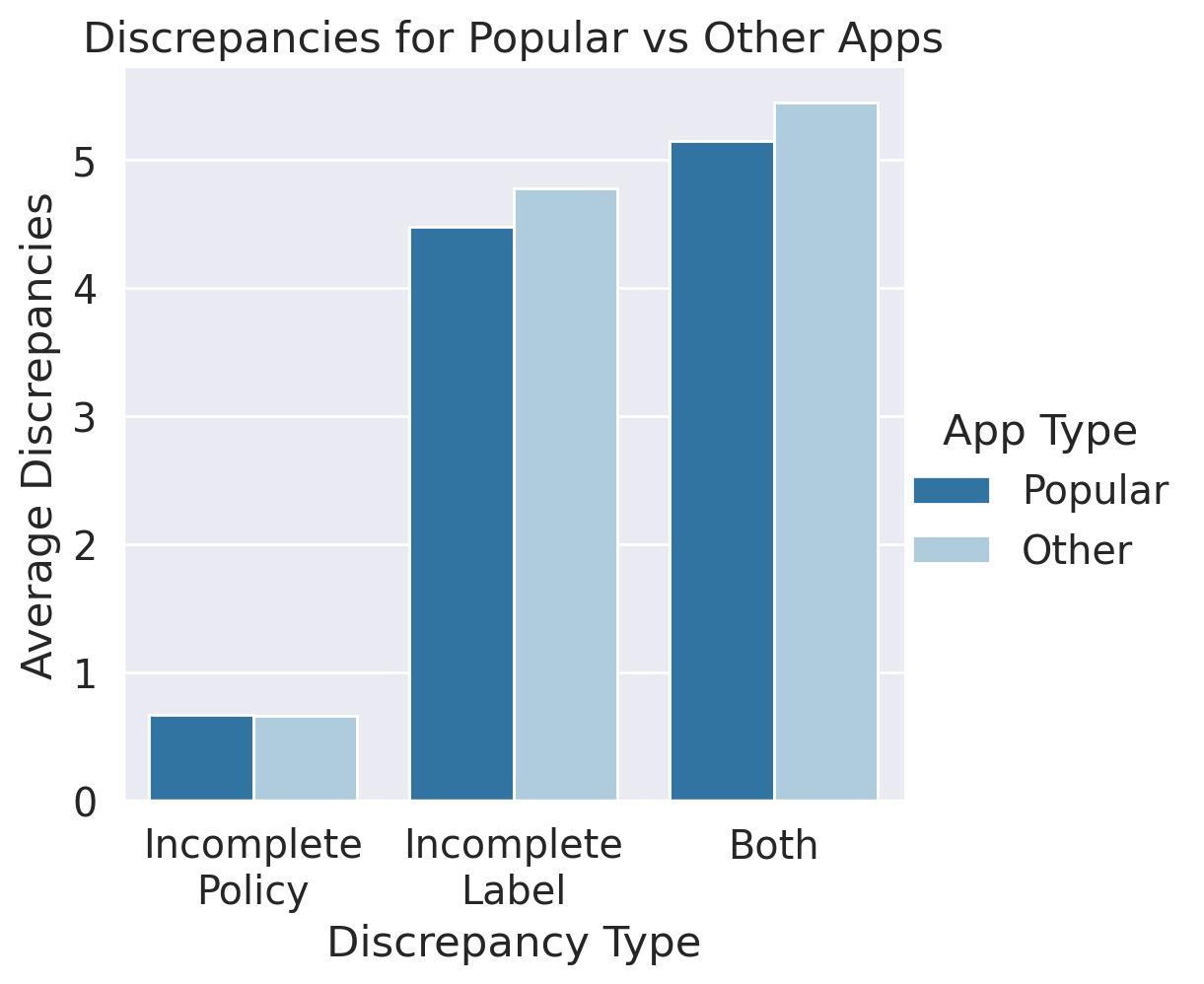}
    \caption[Discrepancies Among Popular vs Other Apps]{A depiction of the average number of discrepancies among popular and other apps. Discrepancies are categorized by Incomplete Policy, Incomplete Label, and Both.}
    \label{fig:popular-vs-other-compliance}
\end{figure}

\begin{table}[h]
    \centering
    \caption[Discrepancies Among Popular vs Other Apps]{This table provides the average number of discrepancies among popular and other apps, categorized by discrepancy type. Discrepancy types with significant difference ($p < 0.05$) are signified with $^{*}$. \label{tab:popular-vs-other-compliance}}
    \footnotesize
    \begin{tabular}{p{0.15\linewidth}p{0.25\linewidth}p{0.25\linewidth}p{0.12\linewidth}}
        \toprule
        {} & \textbf{Popular Apps} & \textbf{Other Apps} & \textbf{p value} \\
        \toprule
        \textbf{Incomplete Policies} & 0.67 & 0.66 & 0.877 \\
        \textbf{Incomplete Labels} &  4.48 & 4.78 & 0.005$^{*}$ \\
        \textbf{Both} &  5.15 & 5.44 & 0.004$^{*}$ \\
        \bottomrule
    \end{tabular}
\end{table}

Finally, we compared the incidence of discrepancies between popular and other apps. As shown in Figure \ref{fig:popular-vs-other-compliance} and Table \ref{tab:popular-vs-other-compliance}, other apps (i.e. unpopular apps), tend to have a statistically significantly higher number of incomplete label discrepancies, on average, than popular apps ($p < 0.05$). Consequently, when looking across all discrepancies, other apps also have a higher number of discrepancies, on average, than popular apps ($p < 0.05$). There is no significant difference between the incidence of incomplete policy discrepancies between popular and other apps ($ p > 0.05$).

We conclude from this comparison that popular apps, on average, have fewer discrepancies than their counterparts. As popular apps represent apps likely to be on user's devices, this trend is reassuring; however, the incidence of discrepancies among popular apps is still high, at an average of 5.15 discrepancies per app.
\section{Discussion}

In Section \ref{chap:3-scaling}, we outlined the development of a pipeline to systematically download and analyze iOS App Store listings. We found 62.5\% of apps to provide privacy labels. This is approximately the same, albeit slightly higher than reported by Balash et al. in their work \cite{balash2022longitudinal}. The slightly higher percentage in our study likely reflects our study being conducted several months after Balash et al.'s, allowing more time for apps to include privacy labels.

In Section \ref{chap:4-generating}, we outlined a technique for predicting iOS privacy labels from the text of privacy policies. To our knowledge, this is the first work that has attempted such a task. However, we acknowledge that our approach has some potential limitations. In certain cases, Apple stipulates that privacy labels may optionally disclose certain data collection, but are not required to do so \cite{apple2020labels}. This could lead to results that overestimate potential discrepancies when looking at optional privacy label disclosures. In principle, because less data collection is typically seen as desirable, developers would typically be expected to only disclose collection when required to do so. So, assuming that our training data is biased towards required disclosures, our resulting classifiers could be expected to indicate positive instances only if they are truly required.

In Section \ref{chap:5-compliance}, we described a large-scale analysis of apps available on the iOS App Store. Prior work had already provided evidence that iOS privacy labels can be inaccurate \cite{li2022understanding}, \cite{zhang2023privacy}. Our study seems to corroborate these earlier results, though there may be multiple possible interpretations of our results. To the extent that discrepancies are indicative of potential compliance issues, we find that as many as 88.0\% of apps have at least one potential compliance issue, with apps having an average of 5.32 potential compliance issues. These results appear generally consistent with prior work, albeit in slightly different contexts. For example, Zimmeck et al. report a mean of 2.89 potential compliance issues per app in their large-scale analysis of Android apps; however, they characterized potential compliance issues as discrepancies between the text of privacy policies and static code analysis of Android apps \cite{zimmeck2019maps}. Our work, on the other hand, compares disclosures within privacy policies to those in privacy labels. We also find that Incomplete Label discrepancies were more common than Incomplete Policy discrepancies. This could likely be a byproduct of privacy policies being written to be more general and permissive than privacy labels. Privacy policies are more likely to be written by lawyers, whereas privacy labels are more likely to be created by developers. Additionally, some policies are known to apply to multiple apps, and even some websites, which could explain why they may disclose practices that an app may not engage in. Privacy labels, on the other hand, are created for each app and would be more likely to only disclose practices an app actually engages in. Accordingly, incomplete privacy policies are more likely to be indicative of privacy compliance issues than incomplete privacy labels.

Several avenues exist to extend this work. For example, future work could extend our approach beyond the identification of data collection disclosures and include disclosures related to how data is used. Additionally, future research could compare the text of iOS privacy policies to static and dynamic analyses of iOS apps, which would provide a more precise understanding of whether the discrepancies reported in our research correspond to actual compliance issues or whether they just reflect different approaches to deciding what to disclose.
\section{Conclusion}

Privacy labels have been proposed as usable mechanisms to help users better understand salient data practices found within privacy policies. In December 2020, Apple began requiring that all iOS apps include privacy labels, arguably the largest adoption of privacy labels to date; however, prior work has questioned the accuracy of such labels \cite{li2022understanding} \cite{gardner2022helping}.

In this work, we introduced the Automated Privacy Label Analysis System (ATLAS). ATLAS enabled us to provide a detailed analysis of 354,725 iOS apps available on the United States iOS App Store. We found that privacy policy accessibility and privacy label adoption is relatively low, with only 62.5\% of apps providing privacy labels. We also developed an ensemble-based classier that was able to accurately predict privacy labels from privacy policies with 91.3\% accuracy. We then used our classifier to conduct a compliance analysis, finding several interesting trends. For example, 88.0\% of apps had at least one discrepancy between the text of their privacy policy and their privacy label. On average, we found iOS apps to have 5.32 discrepancies. These discrepancies could potentially be indicative of compliance issues. We hope that our work enables a thorough review of privacy labels to help promote accurate privacy disclosures in the future.

\section*{Acknowledgements}
This research has been supported in part through grants from the National Science Foundation under its Secure and Trustworthy Computing program (CNS-1801316 and CNS-1914486) and in part through an unrestricted grant from Google under its ``privacy-related faculty award'' program. The US Government is authorized to reproduce and distribute reprints for Government purposes notwithstanding any copyright notice. The views and conclusions contained herein are those of the authors and should not be interpreted as representing the official policies or endorsements, either expressed or implied of NSF, the US Government, or Google. This research has been partially supported by the project TED2021-130455A-I00 funded by MCIN/AEI/10.13039/501100011033 and the Europea Union “NextGenerationEU”/PRTR.

\bibliographystyle{plain}

\bibliography{bibliography}

\onecolumn
\appendices

\section{Data Collection by Data Type}
\begin{figure}[h!]
    \centering
    \includegraphics[height=3in]{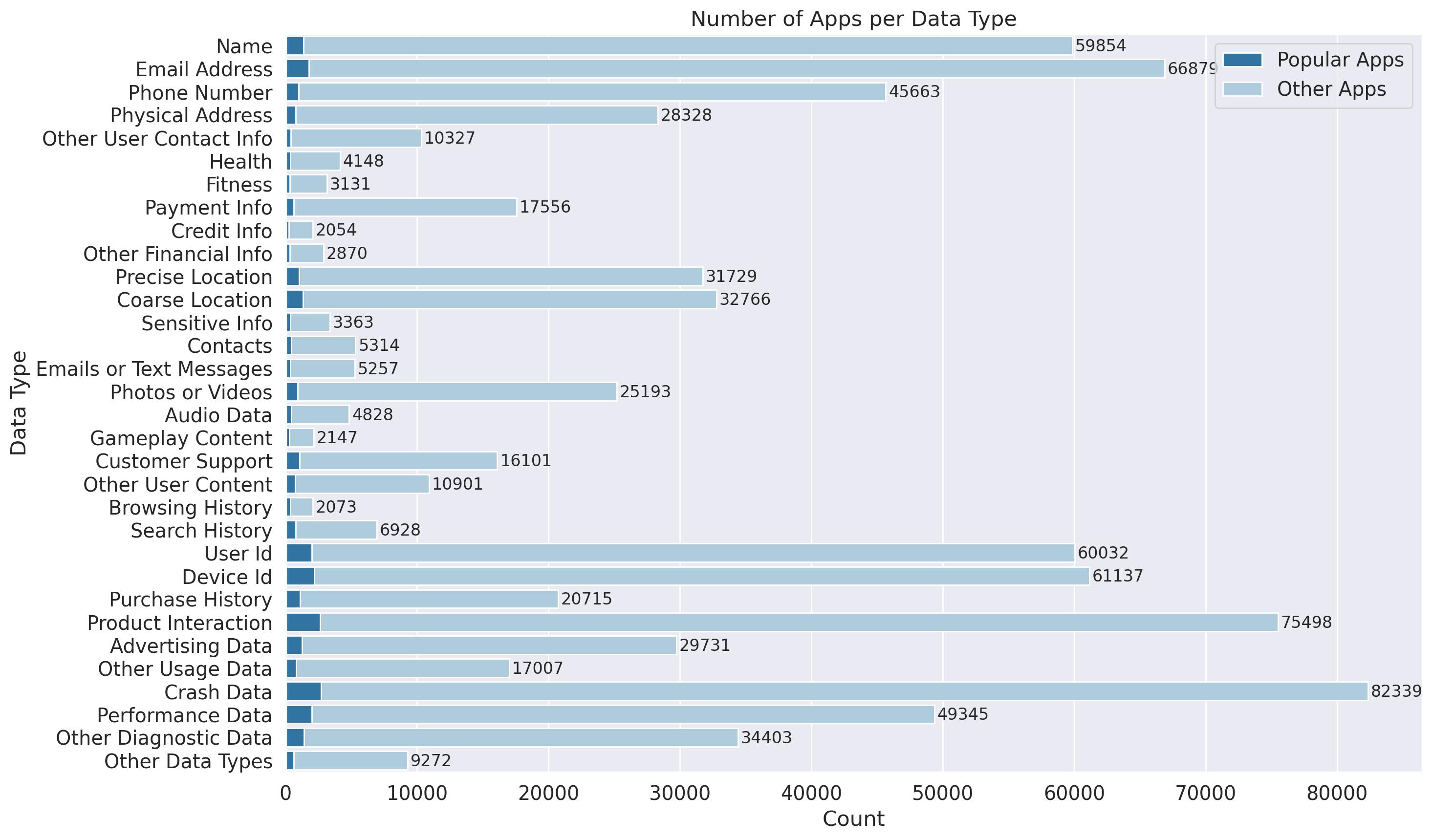}
    \caption[Data Type Declaration by App]{A depiction of the number of apps declaring collection of each data type.}
    \label{fig:apps-per-data-type}
\end{figure}

\newpage
\section{Ensemble Model Test Results}
\begin{table}[h]
    \centering
    \caption[Ensemble Test Performance]{The following table summarizes the performance of the final ensemble model on the test dataset. \label{tab:ensemble-test-perf}}
    \small
    \begin{tabular}{lrrllrr}
        \toprule
        {} & Test Acc (-/+) &  Macro F1 &    F1 (-/+) &  Prec (-/+) & Recall (-/+) & Support (-/+) \\
        \toprule
        Name                    &     0.99, 0.92 &      0.95 &  0.95, 0.95 &  0.93, 0.99 &   0.99, 0.92 &        75, 75 \\
        Email Address           &     0.89, 0.96 &      0.93 &  0.92, 0.93 &  0.96, 0.90 &   0.89, 0.96 &        75, 75 \\
        Phone Number            &     0.91, 0.95 &      0.93 &  0.93, 0.93 &  0.94, 0.91 &   0.91, 0.95 &        75, 75 \\
        Physical Address        &     0.95, 1.00 &      0.97 &  0.97, 0.97 &  1.00, 0.95 &   0.95, 1.00 &        75, 75 \\
        Other User Contact Info &     0.96, 1.00 &      0.98 &  0.98, 0.98 &  1.00, 0.96 &   0.96, 1.00 &        75, 75 \\
        Health                  &     0.92, 0.97 &      0.95 &  0.95, 0.95 &  0.97, 0.92 &   0.92, 0.97 &        75, 75 \\
        Fitness                 &     0.93, 0.92 &      0.93 &  0.93, 0.93 &  0.92, 0.93 &   0.93, 0.92 &        75, 75 \\
        Payment Info            &     0.99, 0.96 &      0.97 &  0.97, 0.97 &  0.96, 0.99 &   0.99, 0.96 &        75, 75 \\
        Credit Info             &     1.00, 1.00 &      1.00 &  1.00, 1.00 &  1.00, 1.00 &   1.00, 1.00 &        75, 75 \\
        Other Financial Info    &     0.97, 1.00 &      0.99 &  0.99, 0.99 &  1.00, 0.97 &   0.97, 1.00 &        75, 75 \\
        Precise Location        &     0.96, 0.99 &      0.97 &  0.97, 0.97 &  0.99, 0.96 &   0.96, 0.99 &        75, 75 \\
        Coarse Location         &     0.79, 0.89 &      0.84 &  0.83, 0.85 &  0.88, 0.81 &   0.79, 0.89 &        75, 75 \\
        Sensitive Info          &     0.89, 0.96 &      0.93 &  0.92, 0.93 &  0.96, 0.90 &   0.89, 0.96 &        75, 75 \\
        Contacts                &     1.00, 0.99 &      0.99 &  0.99, 0.99 &  0.99, 1.00 &   1.00, 0.99 &        75, 75 \\
        Emails or Text Messages &     0.88, 1.00 &      0.94 &  0.94, 0.94 &  1.00, 0.89 &   0.88, 1.00 &        75, 75 \\
        Photos or Videos        &     0.91, 0.95 &      0.93 &  0.93, 0.93 &  0.94, 0.91 &   0.91, 0.95 &        75, 75 \\
        Audio Data              &     0.96, 0.97 &      0.97 &  0.97, 0.97 &  0.97, 0.96 &   0.96, 0.97 &        75, 75 \\
        Gameplay Content        &     0.92, 0.93 &      0.93 &  0.93, 0.93 &  0.93, 0.92 &   0.92, 0.93 &        75, 75 \\
        Customer Support        &     0.92, 0.92 &      0.92 &  0.92, 0.92 &  0.92, 0.92 &   0.92, 0.92 &        75, 75 \\
        Other User Content      &     0.89, 0.99 &      0.94 &  0.94, 0.94 &  0.99, 0.90 &   0.89, 0.99 &        75, 75 \\
        Browsing History        &     0.73, 0.69 &      0.71 &  0.72, 0.71 &  0.71, 0.72 &   0.73, 0.69 &        75, 75 \\
        Search History          &     0.80, 0.89 &      0.85 &  0.84, 0.85 &  0.88, 0.82 &   0.80, 0.89 &        75, 75 \\
        User Id                 &     0.96, 0.97 &      0.97 &  0.97, 0.97 &  0.97, 0.96 &   0.96, 0.97 &        75, 75 \\
        Device Id               &     0.77, 0.93 &      0.85 &  0.84, 0.86 &  0.92, 0.80 &   0.77, 0.93 &        75, 75 \\
        Purchase History        &     0.80, 0.99 &      0.89 &  0.88, 0.90 &  0.98, 0.83 &   0.80, 0.99 &        75, 75 \\
        Product Interaction     &     1.00, 0.99 &      0.99 &  0.99, 0.99 &  0.99, 1.00 &   1.00, 0.99 &        75, 75 \\
        Advertising Data        &     0.88, 0.88 &      0.88 &  0.88, 0.88 &  0.88, 0.88 &   0.88, 0.88 &        75, 75 \\
        Other Usage Data        &     0.67, 0.83 &      0.75 &  0.72, 0.77 &  0.79, 0.71 &   0.67, 0.83 &        75, 75 \\
        Crash Data              &     0.92, 0.96 &      0.94 &  0.94, 0.94 &  0.96, 0.92 &   0.92, 0.96 &        75, 75 \\
        Performance Data        &     0.68, 0.87 &      0.77 &  0.75, 0.79 &  0.84, 0.73 &   0.68, 0.87 &        75, 75 \\
        Other Diagnostic Data   &     0.93, 1.00 &      0.97 &  0.97, 0.97 &  1.00, 0.94 &   0.93, 1.00 &        75, 75 \\
        Other Data Types        &     0.77, 0.63 &      0.70 &  0.72, 0.68 &  0.67, 0.73 &   0.77, 0.63 &        75, 75 \\
        \bottomrule
    \end{tabular}
\end{table}

\newpage
\section{Probability Distributions for Data Types After Classification}
\begin{figure}[h]
    \centering
    \begin{subfigure}{0.49\linewidth}
        \centering
        \includegraphics[height=3in]{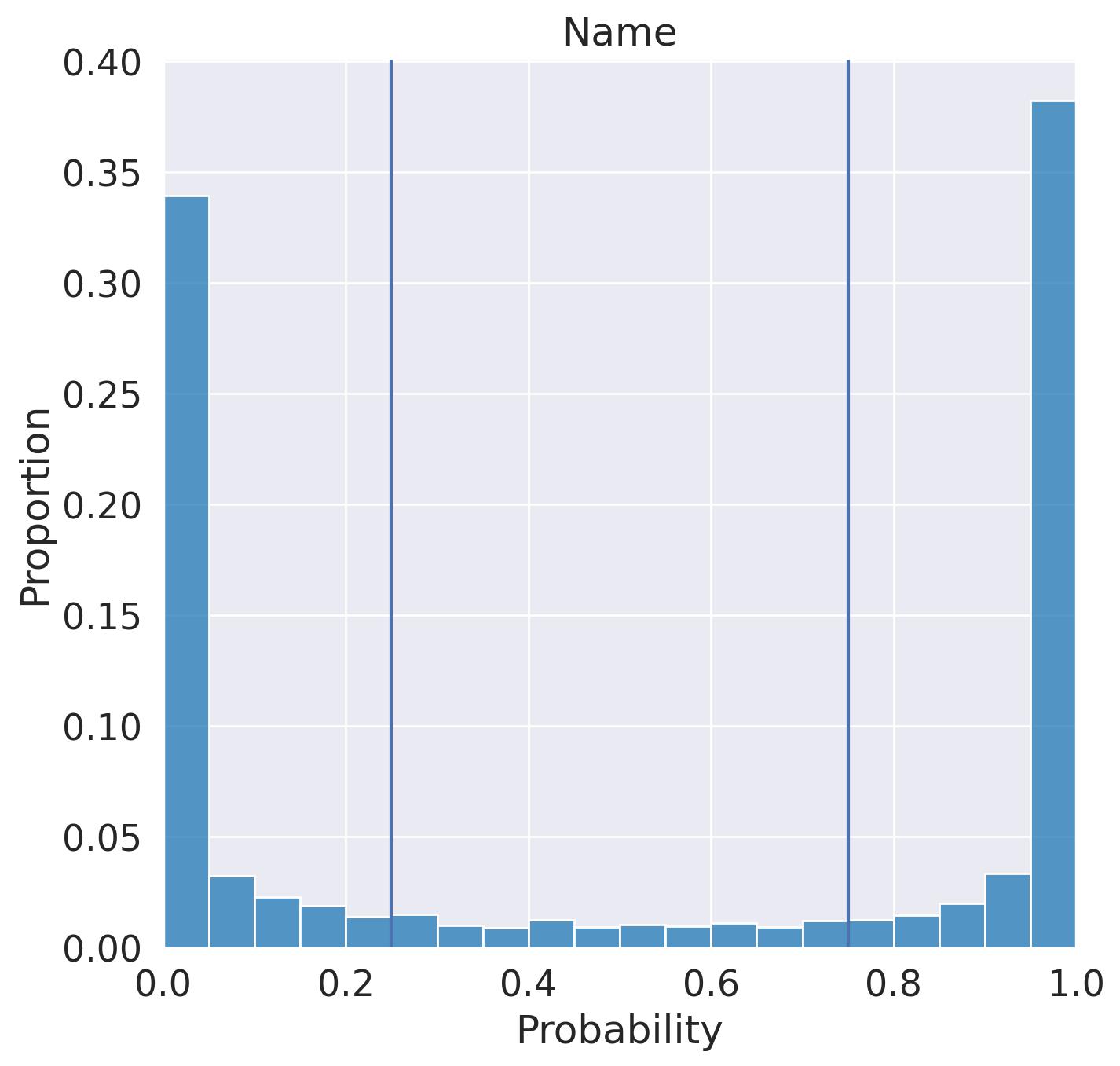}
        \caption{Probability Distribution for Name}
        \label{fig:pdf-name}
    \end{subfigure}
    \begin{subfigure}{0.49\linewidth}
        \centering
        \includegraphics[height=3in]{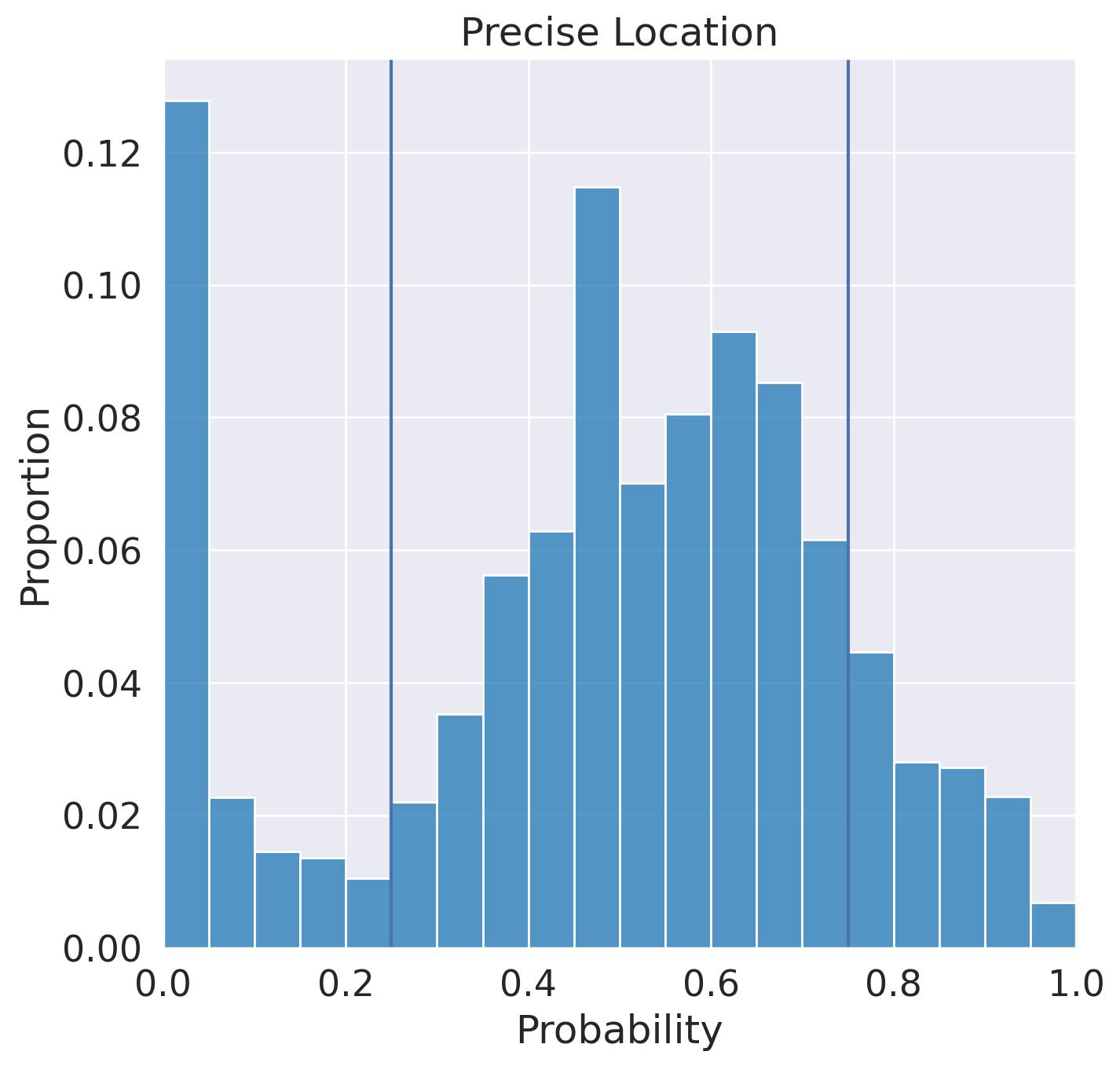}
        \caption{Probability Distribution for Precise Location}
        \label{fig:pdf-precise-location}
    \end{subfigure}
    \caption[Probability Distribution for Name and Precise Location]{Shown here are two probability distributions (Name and Precise Location) after running our ensemble-based classifier on the entire set of privacy policies (excluding training data). This figure characterizes classifier confidence for each data type.}
    \label{fig:pdfs-name-precise-location}
\end{figure}

\newpage
\section{Distribution of Potential Compliance Issues}
\begin{figure}[h]
    \centering
    \begin{subfigure}{0.6\linewidth}
        \centering
        \includegraphics[height=3in]{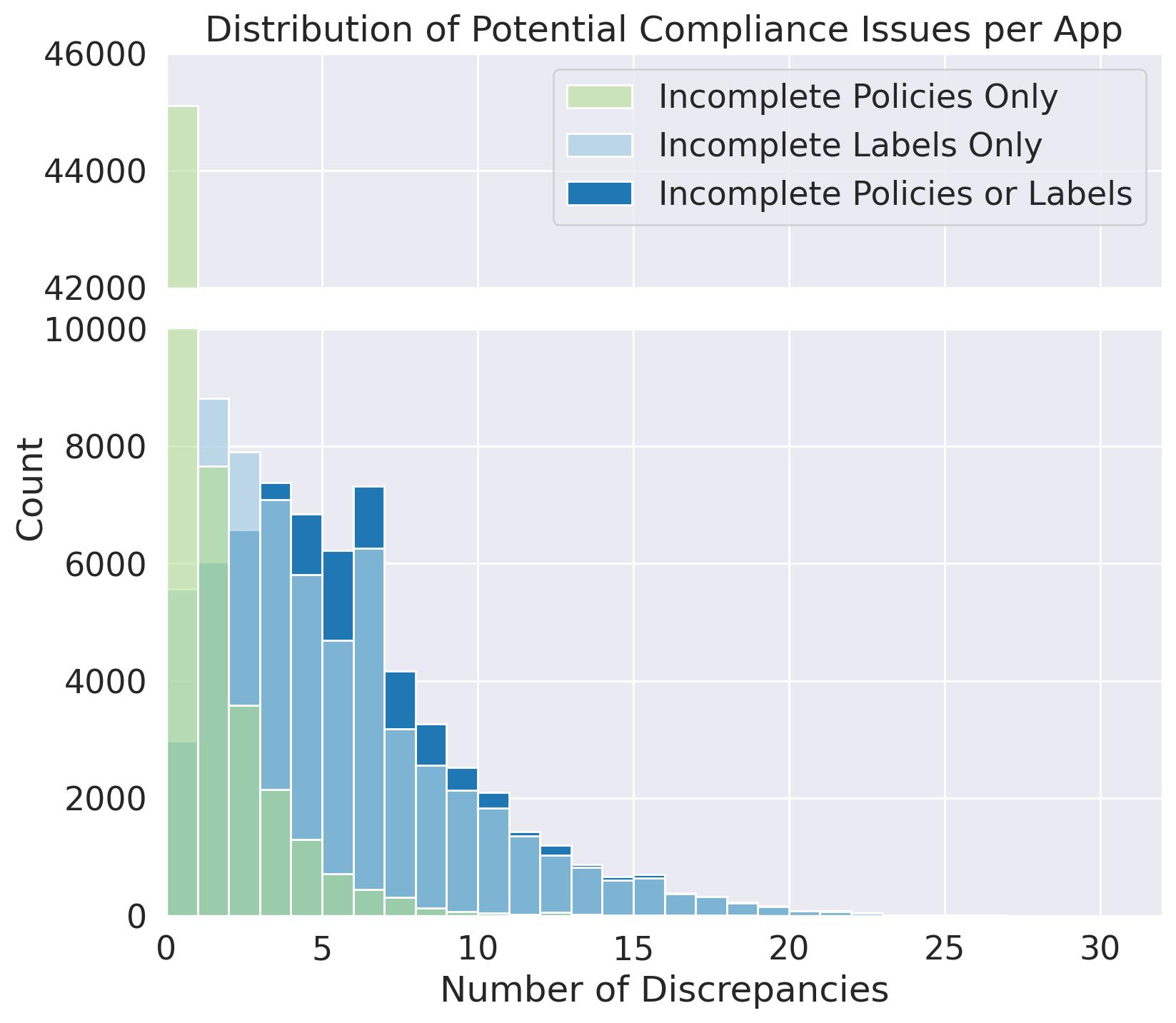}
        \caption{Distribution of Potential Compliance Issues}
        \label{fig:distribution-compliance-issues}
    \end{subfigure}
    \begin{subfigure}{0.38\linewidth}
        \centering
        \includegraphics[height=3in]{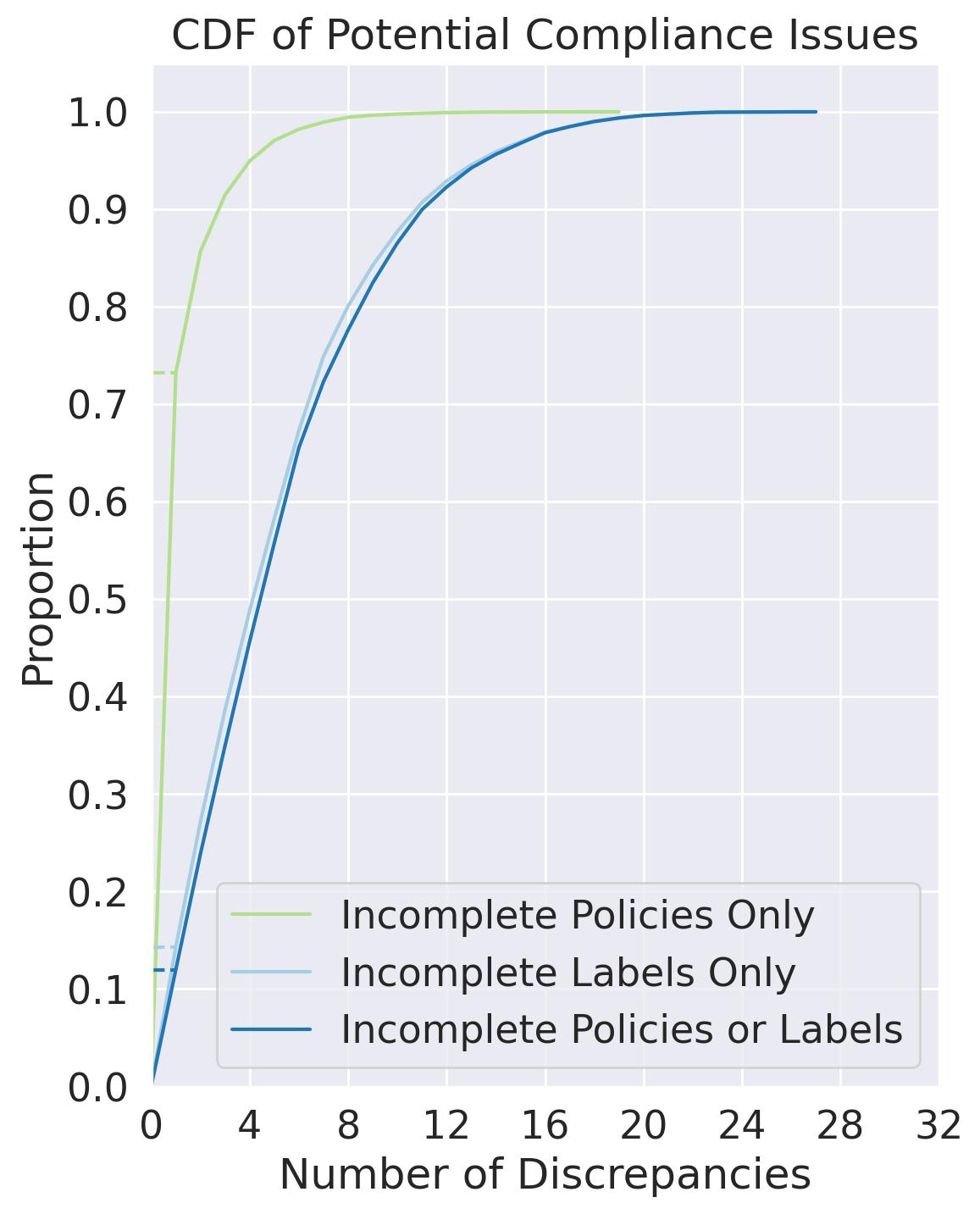}
        \caption{CDF of Potential Compliance Issues}
        \label{fig:cdf-compliance-issues}
    \end{subfigure}
    \caption[Distribution of Compliance Issues]{On the left is a graph of the distribution of the different types of potential compliance issues, and on the right is a CDF of the different types of potential compliance issues.}
    \label{fig:distribution-cdf-compliance-issues}
\end{figure}

\end{document}